\documentclass[12pt]{article}
\usepackage{amsmath}
\usepackage{myart}

\renewcommand{\theequation}{\arabic{section}.\arabic{equation}}
\normalbaselines
\input tcilatex.tex

\begin{document}

\author{Yuri A. Rylov}
\title{Formalized procedure of transition to classical limit in application
to the Dirac equation.}
\date{Institute for Problems in Mechanics, Russian Academy of Sciences \\
101-1 ,Vernadskii Ave., Moscow, 119526, Russia \\
email: rylov@ipmnet.ru\\
Web site: {$http://rsfq1.physics.sunysb.edu/\symbol{126}rylov/yrylov.htm$}\\
or mirror Web site: {$http://195.208.200.111/\symbol{126}rylov/yrylov.htm$}}
\maketitle

\begin{abstract}
Classical model $S_{\mathrm{Dcl}}$ of the Dirac particle $S_{\mathrm{D}}$ is
constructed. $S_{\mathrm{D}}$ is the dynamic system described by the Dirac
equation. For investigation of $S_{\mathrm{D}}$ and construction of $S_{%
\mathrm{Dcl}}$ one uses a new dynamic method: dynamic disquantization. This
relativistic purely dynamic procedure does not use principles of quantum
mechanics. The obtained classical analog $S_{\mathrm{Dcl}}$ is described by
a system of ordinary differential equations, containing the quantum constant
as a parameter. Dynamic equations for $S_{\mathrm{Dcl}}$ are determined by
the Dirac equation uniquely. The dynamic system $S_{\mathrm{Dcl}}$ has ten
degrees of freedom and cannot be a pointlike particle, because it has an
internal structure. Internal degrees of freedom appears to be described
nonrelativistically. One discusses interplay between the conventional
axiomatic methods and the dynamical methods of the quantum systems
investigation. In particular, one discusses the reasons, why the internal
degrees of freedom of the Dirac particle and their nonrelativistic character
were not discovered during eighty years.
\end{abstract}


\section{Introduction}

In the framework of axiomatic presentation of quantum mechanics there exist
no formal procedure of transition to classical approximation. The classical
description is obtained from the quantum one, when we set, that the quantum
constant $\hbar =0$. Unfortunately, we cannot obtain the classical
approximation for the Schr\"{o}dinger particle $\mathcal{S}_{\mathrm{S}}$,
setting $\hbar =0$ in the action for the Schr\"{o}dinger particle. This
action has the form 
\begin{equation}
\mathcal{S}_{\mathrm{S}}:\qquad \mathcal{A}_{\mathrm{S}}\left[ \psi ,\psi
^{\ast }\right] =\int \left\{ \frac{i\hbar }{2}\left( \psi ^{\ast }\partial
_{0}\psi -\partial _{0}\psi ^{\ast }\cdot \psi \right) -\frac{\hbar ^{2}}{2m}%
\mathbf{\nabla }\psi ^{\ast }\mathbf{\nabla }\psi \right\} dtd\mathbf{x}
\label{a1.2}
\end{equation}%
where $\psi =\psi \left( t,\mathbf{x}\right) $ is a complex one-component
wave function, $\psi ^{\ast }=\psi ^{\ast }\left( t,\mathbf{x}\right) $ is
the quantity complex conjugate to $\psi $, and $m$ is the particle mass. The
action (\ref{a1.2}) generates the dynamic equation 
\begin{equation}
i\hbar \partial _{0}\psi =-\frac{\hbar ^{2}}{2m}\mathbf{\nabla }^{2}\psi
\label{a1.2a}
\end{equation}%
The 4-current $j^{k}$ and the energy-momentum tensor $T_{l}^{k}$ are the
canonical quantities associated with the action $\mathcal{A}_{\mathrm{S}}%
\left[ \psi ,\psi ^{\ast }\right] $. They are determined by the relations 
\begin{equation}
j^{k}=\left\{ \rho ,\mathbf{j}\right\} =\frac{i}{\hbar }\left( \frac{%
\partial \mathcal{L}}{\partial \left( \partial _{k}\psi ^{\ast }\right) }%
\psi ^{\ast }-\frac{\partial \mathcal{L}}{\partial \left( \partial _{k}\psi
\right) }\psi \right) =\left\{ \psi ^{\ast }\psi ,-\frac{i\hbar }{2m}\left(
\psi ^{\ast }\mathbf{\nabla }\psi -\mathbf{\nabla }\psi ^{\ast }\cdot \psi
\right) \right\}  \label{a1.3}
\end{equation}%
\begin{equation}
T_{l}^{k}=\frac{\partial \mathcal{L}}{\partial \left( \partial _{k}\psi
^{\ast }\right) }\partial _{l}\psi ^{\ast }+\frac{\partial \mathcal{L}}{%
\partial \left( \partial _{k}\psi \right) }\partial _{l}\psi -\delta _{l}^{k}%
\mathcal{L},\qquad k,l=0,1,2,3  \label{a1.4}
\end{equation}%
where $\mathcal{L}$ is the Lagrangian density for the action (\ref{a1.2}) 
\begin{equation}
\mathcal{L}=\frac{i\hbar }{2}\left( \psi ^{\ast }\partial _{0}\psi -\partial
_{0}\psi ^{\ast }\cdot \psi \right) -\frac{\hbar ^{2}}{2m}\mathbf{\nabla }%
\psi ^{\ast }\mathbf{\nabla }\psi  \label{a1.4a}
\end{equation}

If we set $\hbar =0$ in the action (\ref{a1.2}), the description
degenerates. Such a degeneration is conditioned by the artificial use of the
quantum constant in the action (\ref{a1.2}). The fact is that at the natural
description of the Schr\"{o}dinger particle $\mathcal{S}_{\mathrm{S}} $ the
action contains two independent constants: dynamical constant $b$ and
quantum constant $\hbar $. The dynamical constant $b$ is simply an arbitrary
constant of integration, which can take any nonvanishing value. At such a
natural description the quantum constant $\hbar $ describes the quantum
effects, and quantum description transits to the classical one, if one sets $%
\hbar =0$.

The natural description is obtained from the action (\ref{a1.2}) by means of
the transformation 
\begin{equation}
\psi \rightarrow \Psi _{b}=\left\vert \psi \right\vert \exp \left( \frac{%
\hbar }{b}\log \frac{\psi }{\left\vert \psi \right\vert }\right) ,\qquad
\psi =\left\vert \Psi _{b}\right\vert \exp \left( \frac{b}{\hbar }\log \frac{%
\Psi _{b}}{\left\vert \Psi _{b}\right\vert }\right)  \label{a1.9}
\end{equation}
The natural description is carried out in terms of the wave function $\Psi
_{b}$. The action has the form%
\begin{equation*}
\mathcal{S}_{\mathrm{S}}:\qquad \mathcal{A}_{\mathrm{S}}\left[ \Psi
_{b},\Psi _{b}^{\ast }\right] =\int \left\{ \frac{ib}{2}\left( \Psi
_{b}^{\ast }\partial _{0}\Psi _{b}-\partial _{0}\Psi _{b}^{\ast }\cdot \Psi
_{b}\right) -\frac{b^{2}}{2m}\mathbf{\nabla }\Psi _{b}^{\ast }\mathbf{\nabla 
}\Psi _{b}\right.
\end{equation*}%
\begin{equation}
-\left. \frac{\hbar ^{2}-b^{2}}{2m}\left( \mathbf{\nabla }\left\vert \Psi
_{b}\right\vert \right) ^{2}\right\} dtd\mathbf{x}  \label{a1.10}
\end{equation}

The dynamic equation takes the form 
\begin{equation}
ib\partial _{0}\Psi _{b}=-\frac{b^{2}}{2m}\mathbf{\nabla }^{2}\Psi _{b}-%
\frac{\hbar ^{2}-b^{2}}{8m}\left( \frac{\left( \mathbf{\nabla }\rho \right)
^{2}}{\rho ^{2}}+2\mathbf{\nabla }\frac{\mathbf{\nabla }\rho }{\rho }\right)
\Psi _{b},\qquad \rho \equiv \Psi _{b}^{\ast }\Psi _{b}  \label{a1.11}
\end{equation}

Instead of (\ref{a1.3}), we obtain 
\begin{equation}
\rho =\Psi _{b}^{\ast }\Psi _{b},\qquad \mathbf{j}=-\frac{ib}{2m}\left( \Psi
_{b}^{\ast }\mathbf{\nabla }\Psi _{b}-\mathbf{\nabla }\Psi _{b}^{\ast }\cdot
\Psi _{b}\right)  \label{a1.12}
\end{equation}

We underline that the actions (\ref{a1.2}) and (\ref{a1.10}) describe the
same dynamic system in different dependent variables. The action (\ref{a1.10}%
) contains two parameters $b$ and $\hbar $, whereas the action (\ref{a1.2})
contains only one parameter $\hbar $.

Description in terms of $\Psi _{b}$ is a natural description, because, the
constant $\hbar $ describes the quantum effects, and setting $\hbar =0$ in
the action (\ref{a1.10}), we obtain the action for the statistical ensemble
of free classical particles (this fact has been proved in Appendix A) 
\begin{equation*}
\mathcal{S}_{\mathrm{Scl}}:\qquad \mathcal{A}_{\mathrm{Scl}}\left[ \Psi
_{b},\Psi _{b}^{\ast }\right] =\int \left\{ \frac{ib}{2}\left( \Psi
_{b}^{\ast }\partial _{0}\Psi _{b}-\partial _{0}\Psi _{b}^{\ast }\cdot \Psi
_{b}\right) -\frac{b^{2}}{2m}\mathbf{\nabla }\Psi _{b}^{\ast }\mathbf{\nabla 
}\Psi _{b}\right. 
\end{equation*}%
\begin{equation}
\left. +\frac{b^{2}}{2m}\left( \mathbf{\nabla }\left\vert \Psi
_{b}\right\vert \right) ^{2}\right\} dtd\mathbf{x}  \label{a1.12a}
\end{equation}

If we identify parameters $b$ and $\hbar $ in (\ref{a1.10}) - (\ref{a1.12}),
we obtain the artificial description (\ref{a1.2}) - (\ref{a1.4}), where
setting $\hbar =0$, we set simultaneously $b=0$ and the description
degenerates. At the artificial description the dynamic term and the quantum
term of the dynamic equation are compensated, and the dynamic equation
becomes to be linear. This is the main advantage of the artificial
description.

The Dirac particle is the dynamic system $\mathcal{S}_{\mathrm{D}}$,
described by the Dirac equation. The action $\mathcal{A}_{\mathrm{D}}$ for
the dynamic system $\mathcal{S}_{\mathrm{D}}$ has the form 
\begin{equation}
\mathcal{S}_{\mathrm{D}}:\qquad \mathcal{A}_{\mathrm{D}}[\bar{\psi},\psi
]=c^{2}\int (-mc\bar{\psi}\psi +\frac{i}{2}\hbar \bar{\psi}\gamma
^{l}\partial _{l}\psi -\frac{i}{2}\hbar \partial _{l}\bar{\psi}\gamma
^{l}\psi -\frac{e}{c}A_{l}\bar{\psi}\gamma ^{l}\psi )d^{4}x  \label{b1.1}
\end{equation}%
where $m$ and $e$ are respectively the mass and the charge of the Dirac
particle, and $c$ is the speed of the light. Here $\psi $ is four-component
complex wave function, $\psi ^{\ast }$ is the Hermitian conjugate wave
function, and $\bar{\psi}=\psi ^{\ast }\gamma ^{0}$ is the conjugate one.
The quantgities $\gamma ^{i}$, $i=0,1,2,3$ are $4\times 4$ complex constant
matrices, satisfying the relation 
\begin{equation}
\gamma ^{l}\gamma ^{k}+\gamma ^{k}\gamma ^{l}=2g^{kl}I,\qquad k,l=0,1,2,3.
\label{b1.2}
\end{equation}%
where $I$ is the $4\times 4$ identity matrix, and $g^{kl}=$diag$\left(
c^{-2},-1,-1,-1\right) $ is the metric tensor. The quantity $A_{k}$, $%
k=0,1,2,3$ is the electromagnetic potential. The action (\ref{b1.1})
generates dynamic equation for the dynamic system $\mathcal{S}_{\mathrm{D}}$%
, known as the Dirac equation 
\begin{equation}
\gamma ^{l}\left( -i\hbar \partial _{l}+\frac{e}{c}A_{l}\right) \psi +mc\psi
=0  \label{f1.2}
\end{equation}%
and expressions for physical quantities: the 4-flux $j^{k}$ of particles and
the energy-momentum tensor $T_{l}^{k}$%
\begin{equation}
j^{k}=c^{2}\bar{\psi}\gamma ^{k}\psi ,\qquad T_{l}^{k}=\frac{ic^{2}}{2}%
\left( \bar{\psi}\gamma ^{k}\partial _{l}\psi -\partial _{l}\bar{\psi}\cdot
\gamma ^{k}\psi \right)   \label{f1.3}
\end{equation}

Description of the Dirac particle is also artificial in the sense, that it
is degenerate at $\hbar =0$. Unfortunately, the transformation of the kind (%
\ref{a1.9}) is unknown in the case of the Dirac particle $\mathcal{S}_{%
\mathrm{D}}$, and we are forced to look for another approach to the
derivation of the classical approximation. We use dynamical methods of
investigation. It means the investigation of the dynamic system $\mathcal{S}%
_{\mathrm{D}}$ simply as a dynamic system without a use of quantum
principles. Such an approach admits one to obtain classical approximation of 
$\mathcal{S}_{\mathrm{D}}$ by means of some dynamical procedure, which does
not contain a reference to the quantum constant $\hbar $. This procedure
(dynamical disquantization) is insensitive to the form of application of the
quantum constant (natural or artificial). The dynamical disquantization is a
special case of dynamical methods, applied to the investigation of quantum
systems. We manifest application of dynamical methods in the example of the
Schr\"{o}dinger particle $\mathcal{S}_{\mathrm{S}}$.

\section{Dynamical methods of investigation}

We use the mathematical technique, which is more developed, than
conventional formalism of quantum mechanics. This technique supposes that
all essential information on the quantum dynamical system is contained in
the dynamic system itself. Such specific quantum concepts as the wave
function and principles of quantum mechanics appear to be only the means of
description. The wave function as the means of description may be applied to
both quantum and classical dynamic systems. But the quantum principles may
be applied only to quantum dynamic systems, because they contains some
constraints, which are not satisfied for classical systems. The quantum
system and classical system distinguish dynamically (in additional terms in
the action), but not in the way of description. This fact becomes to be
clear, when both systems are described in the same terms. For instance, the
quantum system and the corresponding classical system may be described in
terms of the wave function, or both systems may be described in terms of the
particle position and momentum.

Progress in the development of the mathematical technique has a \textit{%
mathematical ground: integration of dynamic equations}. This pure
mathematical achievement has physical consequences. It appears that the
quantum mechanics may be considered to be a statistical description of
randomly moving particles. We underline that we investigate well known
quantum systems, and all new results are \textit{corollaries of the more
developed methods of investigation}.

The dynamic system $\mathcal{S}_{\mathrm{S}}$ is determined completely by
dynamic equations (\ref{a1.2a}) and expressions (\ref{a1.3}), (\ref{a1.4})
for the 4-current and the energy-momentum tensor. Only connection between
the particle and the wave function is not described by these relations. This
connection is described by means of the relations

\begin{equation}
\left\langle F\left( \mathbf{x},\mathbf{p}\right) \right\rangle =B\int \func{%
Re}\left\{ \psi ^{\ast }F\left( \mathbf{x},\mathbf{\hat{p}}\right) \psi
\right\} d\mathbf{x,\qquad \hat{p}}=\mathbf{-i\hbar \mathbf{\nabla },\qquad }%
B=\left( \int \psi ^{\ast }\psi d\mathbf{x}\right) ^{-1}  \label{b1.5}
\end{equation}
which define the mean value $\left\langle F\left( \mathbf{x},\mathbf{p}%
\right) \right\rangle $ of any function $F\left( \mathbf{x},\mathbf{p}%
\right) $ of the particle coordinates $\mathbf{x}$ and momentum $\mathbf{p}$%
. Application of the rules (\ref{b1.5}) is restricted by some conditions.
They demand that the dynamic equations be linear and the wave function be a
vector in the Hilbert space of states. We shall refer to the relations (\ref%
{b1.5}) together with the restrictions imposed on its applications as the
quantum principles, because von Neumann has shown \cite{N32}, that the
quantum mechanics can be deduced from relations of the type (\ref{b1.5}),
provided they are valid for all observable quantities. Thus, the
interpretation of the wave function is carried out on the basis of the
quantum principles, which are something external with respect to the dynamic
system $\mathcal{S}_{\mathrm{S}}$.

In reality, the quantum principles are not necessary for interpretation of
the wave function and properties of the dynamic system $\mathcal{S}_{\mathrm{%
S}}$. It is sufficient to make a proper change of dynamic variables and to
describe the dynamic system $\mathcal{S}_{\mathrm{S}}$ in terms of the
particle coordinates $\mathbf{x}$. Such a description does not contain the
enigmatic wave function, whose meaning is unclear, and one does not need the
quantum principles (\ref{b1.5}) for its interpretation. The Schr\"{o}dinger
particle $\mathcal{S}_{\mathrm{S}}$ is a partial case of the generalized Schr%
\"{o}dinger particle $\mathcal{S}_{\mathrm{gS}}$, which is the dynamic
system $\mathcal{S}_{\mathrm{gS}}$, described by the action 
\begin{equation}
\mathcal{A}_{\mathrm{gS}}[\psi ,\psi ^{\ast }]=\int \left\{ \frac{i\hbar }{2}%
\left( \psi ^{\ast }\partial _{0}\psi -\partial _{0}\psi ^{\ast }\cdot \psi
\right) -\frac{\hbar ^{2}}{2m}\mathbf{\nabla }\psi ^{\ast }\mathbf{\nabla }%
\psi +\frac{\hbar ^{2}}{8m}\sum\limits_{\alpha =1}^{\alpha =3}(\mathbf{%
\nabla }s_{\alpha })^{2}\rho \right\} \mathrm{d}^{4}x  \label{b1.5a}
\end{equation}%
\begin{equation}
\rho \equiv \psi ^{\ast }\psi ,\qquad \mathbf{s}\equiv \frac{\psi ^{\ast }%
\mathbf{\sigma }\psi }{\rho },\qquad \mathbf{\sigma }=\{\sigma _{\alpha
}\},\qquad \alpha =1,2,3,  \label{b1.5b}
\end{equation}%
Here $\psi =\left( _{\psi _{2}}^{\psi _{1}}\right) $, $\psi ^{\ast }=\left(
\psi _{1}^{\ast },\psi _{2}^{\ast }\right) $ is the two-component wave
function, and $\sigma _{\alpha }$ are the Pauli matrices. The 4-current is
defined by the relation (\ref{a1.3}) with two-component wave function $\psi $%
. In the case, when components $\psi _{1}$ and $\psi _{2}$ are linear
dependent (for instance, $\psi =\left( _{0}^{\psi _{1}}\right) $), the mean
spin vector $\mathbf{s}=$const, and the last term in the action (\ref{b1.5a}%
) vanishes. In this case the dynamic system $\mathcal{S}_{\mathrm{gS}}$
turns into the dynamic system (\ref{a1.2}).

One can show, that the dynamic system $\mathcal{S}_{\mathrm{gS}}$ is another
representation of the dynamic system $\mathcal{E}\left[ \mathcal{S}_{\mathrm{%
st}}\right] $, i.e. the action for $\mathcal{S}_{\mathrm{gS}}$ can be
obtained from the action for the dynamic system $\mathcal{E}\left[ \mathcal{S%
}_{\mathrm{st}}\right] $ by means of a proper change of variables \cite{R99}.

The dynamic system $\mathcal{E}\left[ \mathcal{S}_{\mathrm{st}}\right] $ is
a statistical ensemble of stochastic particles $\mathcal{S}_{\mathrm{st}}$.
It is described by the action 
\begin{equation}
\mathcal{E}\left[ \mathcal{S}_{\mathrm{st}}\right] :\qquad \mathcal{A}_{%
\mathcal{E}\left[ \mathcal{S}_{\mathrm{st}}\right] }\left[ \mathbf{x,u}_{%
\mathrm{st}}\right] =\int \left\{ \frac{m}{2}\left( \frac{d\mathbf{x}}{dt}%
\right) ^{2}+\frac{m}{2}\mathbf{u}_{\mathrm{st}}^{2}-\frac{\hbar }{2}\mathbf{%
\nabla u}_{\mathrm{st}}\right\} dtd\mathbf{\xi }  \label{a1.22}
\end{equation}%
where $\mathbf{u}_{\mathrm{st}}=\mathbf{u}_{\mathrm{st}}\left( t,\mathbf{x}%
\right) $ is a vector function of arguments $t,\mathbf{x}$ (not of $t,%
\mathbf{\xi }$), and $\mathbf{x}=\mathbf{x}\left( t,\mathbf{\xi }\right) $
is a 3-vector function of independent variables $t,\mathbf{\xi =}\left\{ \xi
_{1,}\xi _{2},\xi _{3}\right\} $. Dynamic equations for the dynamic system $%
\mathcal{E}\left[ \mathcal{S}_{\mathrm{st}}\right] $ are obtained as a
result of variation of the action (\ref{a1.22}) with respect to dependent
dynamic variables $\mathbf{x,u}_{\mathrm{st}}$. In the action (\ref{a1.22})
the variables $\mathbf{\xi }$ label stochastic systems $\mathcal{S}_{\mathrm{%
st}}$, constituting the statistical ensemble. The operator $\mathbf{\nabla }$
is defined in the space of coordinates $\mathbf{x}$ by the relation 
\begin{equation}
\mathbf{\nabla =}\left\{ \partial _{1},\partial _{2},\partial _{3}\right\} 
\mathbf{\equiv }\left\{ \frac{\partial }{\partial x^{1}},\frac{\partial }{%
\partial x^{2}},\frac{\partial }{\partial x^{3}}\right\}  \label{a1.22a}
\end{equation}%
The 3-vector $\mathbf{u}_{\mathrm{st}}$ describes the mean value of the
stochastic component of the particle motion, which is considered to be a
function of the variables $t,\mathbf{x}$. The first term $\frac{m}{2}\left( 
\frac{d\mathbf{x}}{dt}\right) ^{2}$ describes the energy of the regular
component of the stochastic particle motion. The second term $m\mathbf{u}_{%
\mathrm{st}}^{2}/2$ describes the energy of the random component of
velocity. The components $\frac{d\mathbf{x}}{dt}$ and $\mathbf{u}_{\mathrm{st%
}}$ of the total velocity are connected with different degrees of freedom,
and their energies should be added in the expression for the Lagrange
function density. The last term $-\hbar \mathbf{\nabla u}_{\mathrm{st}}/2$
describes interplay between the velocity $\frac{d\mathbf{x}}{dt}$ of the
regular component and the random one $\mathbf{u}_{\mathrm{st}}$.

The action (\ref{a1.22}) is a sum (integral) of actions for independent
stochastic systems $\mathcal{S}_{\mathrm{st}}$, labelled by the parameters $%
\mathbf{\xi }=\left\{ \xi _{1},\xi _{2},\xi _{3}\right\} $. Any stochastic
system $\mathcal{S}_{\mathrm{st}}$ is a stochastic particle, whose state is
described by its coordinate $\mathbf{x}\left( t\right) $. The action for the
stochastic system $\mathcal{S}_{\mathrm{st}}$ is obtained from the action (%
\ref{a1.22}) for $\mathcal{E}\left[ \mathcal{S}_{\mathrm{st}}\right] $. It
has the form 
\begin{equation}
\mathcal{S}_{\mathrm{st}}:\qquad \mathcal{A}_{\mathcal{S}_{\mathrm{st}}}%
\left[ \mathbf{x,u}_{\mathrm{st}}\right] =\int \left\{ \frac{m}{2}\left( 
\frac{d\mathbf{x}}{dt}\right) ^{2}+\frac{m}{2}\mathbf{u}_{\mathrm{st}}^{2}-%
\frac{\hbar }{2}\mathbf{\nabla u}_{\mathrm{st}}\right\} dt  \label{a1.23}
\end{equation}%
where $\mathbf{x}=\mathbf{x}\left( t\right) $. In reality, the action (\ref%
{a1.23}) is not well defined mathematically, if $\hbar \neq 0$. It is only
symbolic, because the operator (\ref{a1.22a}) is defined in the vicinity of
the point $\mathbf{x}$, but not at the point $\mathbf{x}$ itself. As a
result the dynamic equations for the stochastic system $\mathcal{S}_{\mathrm{%
st}}$ do not exist, if $\hbar \neq 0$. This fact agrees with the
stochasticity of $\mathcal{S}_{\mathrm{st}}$. By definition the system $%
\mathcal{S}_{\mathrm{st}}$ is stochastic, if there exist no dynamic
equations for $\mathcal{S}_{\mathrm{st}}$. If we cut off interaction with
the stochastic agent, setting $\hbar =0$ in the action (\ref{a1.23}) (or$\ $%
remove two last terms), we obtain the well defined action for the free
nonrelativistic deterministic particle $\mathcal{S}_{\mathrm{d}}$%
\begin{equation}
\mathcal{S}_{\mathrm{d}}:\qquad \mathcal{A}_{\mathcal{S}_{\mathrm{d}}}\left[ 
\mathbf{x,u}_{\mathrm{st}}\right] =\int \left\{ \frac{m}{2}\left( \frac{d%
\mathbf{x}}{dt}\right) ^{2}+\frac{m}{2}\mathbf{u}_{\mathrm{st}}^{2}\right\}
dt,\qquad \mathbf{x}=\mathbf{x}\left( t\right)  \label{a1.23a}
\end{equation}

The Schr\"{o}dinger particle $\mathcal{S}_{\mathrm{S}}$ (\ref{a1.2}) is a
partial case of the dynamic system $\mathcal{E}\left[ \mathcal{S}_{\mathrm{st%
}}\right] $ (\ref{a1.22}), whereas the generalized Schr\"{o}dinger particle $%
\mathcal{S}_{\mathrm{gS}}$ (\ref{b1.5a}) coincide with the dynamic system $%
\mathcal{E}\left[ \mathcal{S}_{\mathrm{st}}\right] $ (\ref{a1.22}). The
action (\ref{b1.5a}) may be obtained from the action (\ref{a1.22})
mathematically by means of a proper change of variables. (see Appendix A).

Interpretation of the dynamic system (\ref{a1.22}) is very simple, but
dynamic equations for $\mathcal{E}\left[ \mathcal{S}_{\mathrm{st}}\right] $
are rather complicated. They have the form 
\begin{equation}
\frac{\delta \mathcal{A}_{\mathcal{E}\left[ \mathcal{S}_{\mathrm{st}}\right]
}}{\delta \mathbf{x}}=-m\frac{d^{2}\mathbf{x}}{dt^{2}}+\mathbf{\nabla }%
\left( \frac{m}{2}\mathbf{u}_{\mathrm{st}}^{2}-\frac{\hbar }{2}\mathbf{%
\nabla u}_{\mathrm{st}}\right) =0  \label{b1.4}
\end{equation}%
\begin{equation}
\frac{\delta \mathcal{A}_{\mathcal{E}\left[ \mathcal{S}_{\mathrm{st}}\right]
}}{\delta \mathbf{u}_{\mathrm{st}}}=m\mathbf{u}_{\mathrm{st}}\rho +\frac{%
\hbar }{2}\mathbf{\nabla }\rho =0,  \label{b1.7}
\end{equation}%
where $\rho $ is the function of derivatives of $\mathbf{x}$ with respect to 
$\mathbf{\xi }=\left\{ \xi _{1},\xi _{2},\xi _{3}\right\} $, determined by
the relation 
\begin{equation}
\rho =\left[ \frac{\partial \left( x^{1},x^{2},x^{3}\right) }{\partial
\left( \xi _{1},\xi _{2},\xi _{3}\right) }\right] ^{-1}=\frac{\partial
\left( \xi _{1},\xi _{2},\xi _{3}\right) }{\partial \left(
x^{1},x^{2},x^{3}\right) }  \label{b1.9}
\end{equation}%
Resolving the relation (\ref{b1.7}) with respect to $\mathbf{u}_{\mathrm{st}%
} $ in the form 
\begin{equation}
\mathbf{u}_{\mathrm{st}}=-\frac{\hbar }{2m}\mathbf{\nabla }\ln \rho ,
\label{b1.10}
\end{equation}%
and eliminating $\mathbf{u}_{\mathrm{st}}$ from (\ref{b1.4}), we obtain 
\begin{equation}
m\frac{d^{2}\mathbf{x}}{dt^{2}}=-\mathbf{\nabla }U\left( \rho ,\mathbf{%
\nabla }\rho \right) ,\qquad U\left( \rho ,\mathbf{\nabla }\rho \right) =%
\frac{\hbar ^{2}}{8m}\left( \frac{\left( \mathbf{\nabla }\rho \right) ^{2}}{%
\rho ^{2}}-2\frac{\mathbf{\nabla }^{2}\rho }{\rho }\right)  \label{b1.11}
\end{equation}%
Thus, dynamic equations, generated by the action (\ref{a1.22}), describe the
regular motion component of any particle $\mathcal{S}_{\mathrm{st}}$, as a
motion in a very complicated potential field $U$, depending on the
distribution of all particles of the statistical ensemble $\mathcal{E}\left[ 
\mathcal{S}_{\mathrm{st}}\right] $. Of course, the trajectories $\mathbf{x}=%
\mathbf{x}\left( t,\mathbf{\xi }\right) $ do not describe the motion of
individual stochastic particles. They describe only statistical average
motion of stochastic particles. The situation reminds situation in the gas
dynamics. The dynamic equations of the gas dynamics describe the motion of
the "gas particles", which contain many molecules. Motion of the gas
molecules is random and chaotic. It cannot be described by the gas dynamics
equations, which describe only regular component of the molecule motion.

Note, that the term $\frac{m}{2}\mathbf{u}_{\mathrm{st}}^{2}$ in (\ref{a1.23}%
) looks as a kinetic energy, but according to (\ref{b1.10}) it does not
depend on the temporary derivative $\mathbf{\dot{x}}$, and in dynamic
equations it acts as a potential energy.

The statistical ensemble (\ref{a1.22}) may be considered to be some fluid.
We may speak about the flow of the statistical ensemble $\mathcal{E}\left[ 
\mathcal{S}_{\mathrm{st}}\right] $, keeping in mind, that dynamic equation (%
\ref{b1.11}) for the dynamic system $\mathcal{E}\left[ \mathcal{S}_{\mathrm{%
st}}\right] $ may be interpreted as hydrodynamic equation for some "quantum"
fluid.

On the contrary, the dynamic equations, generated by the action (\ref{a1.2}%
), are linear and rather simple, whereas their interpretation is very
complicated, because it uses the principles of quantum mechanics (\ref{b1.5}%
). Thus, the description by means of the action (\ref{a1.22}) admits a
simple interpretation, but dynamic equations are very complicated for a
solution.

If the action (\ref{a1.2}) is a special case of the action (\ref{a1.22}), it
is reasonable to use the dynamic system $\mathcal{E}\left[ \mathcal{S}_{%
\mathrm{st}}\right] $ as starting point for the statement of the problem and
for interpretation of the results obtained, whereas the dynamic system $%
\mathcal{S}_{\mathrm{S}}$ will be used only for solution of dynamic
equations, which have a simple form in terms of the wave function. Note,
that according (\ref{s5.32}) the action (\ref{a1.22}) is transformed to the
action (\ref{a1.10}), containing two parameters: arbitrary dynamical
constant $b$ and quantum constant $\hbar $. It is a natural form of the
action, generating nonlinear dynamic equation, if $b\neq \hbar $. Artificial
form (\ref{a1.2}) of the action is obtained after artificial identification
of the dynamical constant with the quantum one.

Why was the statistical ensemble $\mathcal{E}\left[ \mathcal{S}_{\mathrm{st}}%
\right] $ as a starting point was not not used before? Why was the problem
of the stochastic motion of microparticles stated in terms of enigmatic wave
function? The answer is very simple. The connection between two different
forms (\ref{a1.2}) and (\ref{a1.22}) of the action for the Schr\"{o}dinger
particle \textit{has not been known for a long time. }

It is known, that the Schr\"{o}dinger equation can be written in the
hydrodynamical form \cite{M26}. D. Bohm \cite{B52} used this circumstance
for the hydrodynamic interpretation of quantum mechanics. But it was only
interpretation of the quantum principles in the hydrodynamical terms. He
failed to eliminate the quantum principles and the wave function from the
foundation of the quantum mechanics, and the wave function remained to be an
enigmatic object -- the vector in the Hilbert space. One failed to connect
the wave function with the hydrodynamic variables: the density $\rho $ and
the velocity $\mathbf{v}$. In more exact terms the connection between the
wave function and hydrodynamic variables $\rho $, $\mathbf{v}$ was
established, but it was a one-way connection. In the case of the
irrotational flow the hydrodynamical variables can be expressed via the wave
function $\psi $, but one cannot do this in the case of the irrotational
flow. Hence, one can transit from the description in terms of the wave
function to the description in terms of $\rho $, $\mathbf{v}$, but one
cannot transit from the hydrodynamic description in terms of $\rho $, $%
\mathbf{v}$ to a description in terms of $\psi $, because, in general, the
fluid flow is rotational, and the dynamic system (\ref{a1.22}) cannot be
described in terms of the one-component wave function.

Let us present the wave function in the form 
\begin{equation}
\psi =\sqrt{\rho }e^{i\varphi },  \label{b1.14}
\end{equation}%
substitute it in the Schr\"{o}dinger equation (\ref{a1.2a}) and separate the
real and imaginary parts of the equation. We obtain two real equations 
\begin{equation}
\partial _{0}\ln \rho =-\frac{\hbar }{m}\left( \nabla ^{2}\varphi +\nabla
\ln \rho \nabla \varphi \right)  \label{b1.15}
\end{equation}%
\begin{equation}
\partial _{0}\varphi +\frac{\hbar }{2m}\left( \nabla \varphi \right) ^{2}=%
\frac{\hbar }{2m}\left( \frac{1}{2}\nabla ^{2}\ln \rho +\left( \frac{1}{2}%
\nabla \ln \rho \right) ^{2}\right)  \label{b1.16}
\end{equation}%
To obtain hydrodynamic equations, one needs to take gradient of the equation
(\ref{b1.16}) and introduce the velocity $\mathbf{v=}\left\{
v^{1},v^{2},v^{3}\right\} $ by means of the relation 
\begin{equation}
\mathbf{v}=\frac{\hbar }{m}\mathbf{\nabla }\varphi  \label{b1.17}
\end{equation}%
We obtain 
\begin{equation}
\partial _{0}\rho +\partial _{\alpha }\left( \rho v^{\alpha }\right)
=0,\qquad \partial _{0}v^{\alpha }+v^{\beta }\partial _{\beta }v^{\alpha }%
\mathbf{=-}\frac{1}{\rho }\partial _{\beta }P^{\alpha \beta },\qquad \alpha
=1,2,3  \label{b1.18}
\end{equation}%
where $P^{\alpha \beta }$ is the tension tensor 
\begin{equation}
P^{\alpha \beta }=\frac{\hbar ^{2}}{4m^{2}}\left( \frac{\left( \partial
_{\alpha }\rho \right) \partial _{\beta }\rho }{\rho }-\partial _{\alpha
}\partial _{\beta }\rho \right)  \label{b1.18a}
\end{equation}

The hydrodynamic equations (\ref{b1.18}) are obtained as a result of
differentiation of the equation (\ref{a1.2a}), written in terms of the wave
function. It means that to transit from the hydrodynamic equations (\ref%
{b1.18}) to the equation, written in terms of the wave function, one needs
to integrate the hydrodynamic equations (\ref{b1.18}). Besides, in the case
of the irrotational flow the wave function is presented in terms of $\rho $
and hydrodynamical potential $\varphi $. The same is valid in the general
case, but the number of the hydrodynamical potentials is to be more than
one, and it is necessary to introduce additional hydrodynamic (Clebsch)
potentials.

The problem of integration of the hydrodynamical equations is rather
complicated problem, which has been solved only in the end of eighties \cite%
{R89}. To solve this problem, it was necessary to develop a special Jacobian
technique \cite{R99}, which was used already by Clebsch \cite{C57,C59}.

As soon as the hydrodynamic equations for the ideal fluid have been
integrated, it becomes clear, that the wave function is simply a method of
the ideal fluid description. The wave function $\psi $ ceases to be an
enigmatic vector of the Hilbert space, whose meaning was obtained only via
quantum principles. Now one can determine the chain of the dynamic variable
transformations which turn the action (\ref{a1.22}) into the action (\ref%
{b1.5a}) (for details see Appendix A). As a result the action (\ref{a1.22})
may be used as a starting point for the description of the quantum Schr\"{o}%
dinger particle $\mathcal{S}_{\mathrm{S}}$. At such a description the
quantum principles (\ref{b1.5}) are not needed, because they are only a tool
for the interpretation of the wave function.

The statistical ensemble (\ref{a1.22}) as the starting point of the quantum
description has a series of advantages over the action (\ref{a1.2}):

\begin{enumerate}
\item The statistical ensemble (\ref{a1.22}) is a very transparent
construction founded on the simple physical idea, that the quantum particle
is a stochastically moving particle.

\item It does not use quantum principles, which are nonrelativistic and
cannot be extended properly to the relativistical case.

\item Statistical ensemble (\ref{a1.22}) is a more general construction,
because the action (\ref{a1.2}) is a partial case of the action (\ref{a1.22})

\item Description in terms of the dynamic system (\ref{a1.22}) is a
statistical description. As any statistical description it contains two
objects: the individual stochastic particle $\mathcal{S}_{\mathrm{st}}$ and
the statistical average particle $\left\langle \mathcal{S}_{\mathrm{st}%
}\right\rangle $. Respectively there are two kinds of measurements:
individual measurement (S-measurement) produced over the individual particle 
$\mathcal{S}_{\mathrm{st}}$ and the massive measurement (M-measurement)
produced over the statistical average particle $\left\langle \mathcal{S}_{%
\mathrm{st}}\right\rangle $. These measurements have different properties,
and their identification is inadmissible.
\end{enumerate}

The complexity of dynamic equations (\ref{b1.11}) is the only defect of the
statistical description (\ref{a1.22}).

We underline that the transition from the action (\ref{a1.22}) as a starting
point to the action (\ref{a1.2}) is motivated mathematically. No additional
physical arguments have been used for the substantiation of the statistical
ensemble (\ref{a1.22}) as a starting point of the quantum description.

The statistical description, founded on the action (\ref{a1.22}) leads to
the statement that wave function describes a state of the statistical
ensemble $\mathcal{E}\left[ \mathcal{S}_{\mathrm{st}}\right] $, but not a
state of a single quantum particle. Discussion of the question, what object
is described by the wave function, has a long history. Some researchers \cite%
{H55} believe, that the wave function describes the state of a single
quantum particle, whereas other ones \cite{B76,B70} believe that the wave
function describes the state of the statistical ensemble. There is a long
list of different opinions about this question, but we do not present them,
because this problem is not a question of a belief. It can and must be
solved on the basis of the mathematical formalism.

The problem is set as follows. What dynamic system is described by the
action (\ref{a1.2})? A single quantum particle, or a statistical ensemble of
single particles? Let us go to the limit $\hbar \rightarrow 0$. Then the
action (\ref{a1.2}) will describe the classical dynamic system $\mathcal{S}_{%
\mathrm{Scl}}$. If the dynamic system $\mathcal{S}_{\mathrm{Scl}}$ is a
single classical particle, then the wave function describes the state of a
single particle. If the dynamic system $\mathcal{S}_{\mathrm{cl}}$ is a
statistical ensemble of classical particles, then the wave function
describes the state of a statistical ensemble of single particles.

Setting $\hbar =0$ in (\ref{a1.10}), (\ref{a1.11}), we obtain 
\begin{equation*}
\mathcal{S}_{\mathrm{Scl}}:\qquad \mathcal{A}_{\mathrm{Scl}}\left[ \Psi
_{b},\Psi _{b}^{\ast }\right] =\int \left\{ \frac{ib}{2}\left( \Psi
_{b}^{\ast }\partial _{0}\Psi _{b}-\partial _{0}\Psi _{b}^{\ast }\cdot \Psi
_{b}\right) -\frac{b^{2}}{2m}\mathbf{\nabla }\Psi _{b}^{\ast }\mathbf{\nabla 
}\Psi _{b}\right.
\end{equation*}
\begin{equation}
+\left. \frac{b^{2}}{2m}\left( \mathbf{\nabla }\left\vert \Psi
_{b}\right\vert \right) ^{2}\right\} dtd\mathbf{x}  \label{b1.19}
\end{equation}
\begin{equation}
ib\partial _{0}\Psi _{b}=-\frac{b^{2}}{2m}\mathbf{\nabla }^{2}\Psi _{b}+%
\frac{b^{2}}{8m}\left( \frac{\left( \mathbf{\nabla }\rho \right) ^{2}}{\rho
^{2}}+2\mathbf{\nabla }\frac{\mathbf{\nabla }\rho }{\rho }\right) \Psi
_{b},\qquad \rho \equiv \Psi _{b}^{\ast }\Psi _{b}  \label{b1.20}
\end{equation}

The action (\ref{b1.19}) describes the statistical ensemble of free
classical particles and, hence, the wave function describes the statistical
ensemble, but not a single particle. The action (\ref{b1.19}) may not
describe a single classical particle, because the dynamic system (\ref{b1.19}%
) has infinite number of the freedom degrees. As far as the description (\ref%
{b1.19}) in terms of the wave function $\Psi _{b}$ is a limit $\hbar
\rightarrow 0$ of the description in terms of the wave function $\psi $, the
wave function $\psi $ in (\ref{a1.2}) may not describe a single quantum
particle. Thus, the \textit{supposition that the wave function describes a
state of a single particle is incompatible with the quantum mechanics
formalism}.

According to the Copenhagen interpretation of quantum mechanics the wave
function $\psi $ describes the state of a single quantum particle, whereas
the state of a classical particle is described by its position $\mathbf{x}$
and its momentum $\mathbf{p}$. It is supposed that the wave function is a
specific quantum quantity, which has no classical analog. In accordance with
this approach one may not go to the limit $\hbar \rightarrow 0$ in the
action (\ref{a1.2}), because the action vanishes, and the description
degenerates.

The transformation (\ref{a1.9}) changes only the scale of the wave function
phase ln$\left( \psi /\left\vert \psi \right\vert \right) $, and this change
may be very slight. The wave function $\Psi _{b}$ is the valid wave
function, which can be used, in particular, for calculation of average
values by means of the relation (\ref{b1.5}). This calculation may be
produced for any value of the constant $b$. At different values of the
parameter $b$ the wave function $\Psi _{b}$ may describe the same state of $%
\mathcal{S}_{\mathrm{S}}$ , because the state of the dynamic system does not
determine the wave function uniquely, and the same state of $\mathcal{S}_{%
\mathrm{S}}$ may be described by different wave functions. From viewpoint of
the statistical description (\ref{a1.22}) the wave function is not uniquely
defined, because it is constructed of hydrodynamic potentials, i.e. it is a
result of integration of uniquely defined velocity $\mathbf{v}$. The
parameter $b$ in the transformation (\ref{b1.5}) is a constant of
integration.

We may set $b=\hbar $ in the relations (\ref{b1.19}), (\ref{b1.20}) and
obtain a description of "classical particle " in the form containing the
quantum constant $\hbar $. 
\begin{equation*}
\mathcal{S}_{\mathrm{Scl}}:\qquad \mathcal{A}_{\mathrm{Scl}}\left[ \psi
,\psi ^{\ast }\right] =\int \left\{ \frac{i\hbar }{2}\left( \psi ^{\ast
}\partial _{0}\psi -\partial _{0}\psi ^{\ast }\cdot \psi \right) -\frac{%
\hbar ^{2}}{2m}\mathbf{\nabla }\psi ^{\ast }\mathbf{\nabla }\psi \right.
\end{equation*}
\begin{equation}
+\left. \frac{\hbar ^{2}}{2m}\left( \mathbf{\nabla }\left\vert \psi
\right\vert \right) ^{2}\right\} dtd\mathbf{x}  \label{b1.21}
\end{equation}
\begin{equation}
i\hbar \partial _{0}\psi =-\frac{\hbar ^{2}}{2m}\mathbf{\nabla }^{2}\psi +%
\frac{\hbar ^{2}}{8m}\left( \frac{\left( \mathbf{\nabla }\rho \right) ^{2}}{%
\rho ^{2}}+2\mathbf{\nabla }\frac{\mathbf{\nabla }\rho }{\rho }\right) \psi
,\qquad \rho \equiv \psi ^{\ast }\psi  \label{b1.22}
\end{equation}
The same result may be obtained from (\ref{b1.19}), (\ref{b1.20}) by means
of the transformation inverse to the transformation (\ref{a1.9}). Formally
the action (\ref{b1.21}) distinguishes from the action (\ref{a1.2}) in the
last term, which describes a lack of quantum effects. The quantum constant
in two first terms has no relation to quantum effects. The dependence on $%
\hbar $ is conditioned by a special choice of the arbitrary constant $b$.

The action (\ref{b1.21}) describes the dynamic system $\mathcal{S}_{\mathrm{%
Scl}}=\mathcal{E}\left[ \mathcal{S}_{\mathrm{d}}\right] $ in the "quantum
language", i.e. in terms of the wave function. The action 
\begin{equation}
\mathcal{S}_{\mathrm{Scl}}=\mathcal{E}\left[ \mathcal{S}_{\mathrm{d}}\right]
:\qquad \mathcal{A}_{\mathcal{E}\left[ \mathcal{S}_{\mathrm{d}}\right] }%
\left[ \mathbf{x}\right] =\int \frac{m}{2}\left( \frac{d\mathbf{x}}{dt}%
\right) ^{2}dtd\mathbf{\xi }  \label{b1.23}
\end{equation}
where $\mathbf{x}=\mathbf{x}\left( t,\mathbf{\xi }\right) $, describes the
same dynamic system in the "classical language", i.e. in terms of classical
variables $\mathbf{x},\mathbf{p}$. In the same way the action (\ref{a1.2})
describes the dynamic system $\mathcal{S}_{\mathrm{S}}=\mathcal{E}\left[ 
\mathcal{S}_{\mathrm{st}}\right] $ in quantum language, whereas the action (%
\ref{a1.22}) describes the same dynamic system in the classical language. It
is reasonable that the quantum system $\mathcal{S}_{\mathrm{S}}$ is
described simpler in the quantum language, whereas the classical system $%
\mathcal{S}_{\mathrm{Scl}}=\mathcal{E}\left[ \mathcal{S}_{\mathrm{d}}\right] 
$ is described simpler in the classical language. However, it is not a
reason for the statement that the quantum system is to be described in the
quantum language (in terms of the wave function).

Two different description of the classical system $\mathcal{S}_{\mathrm{cl}}$
can be used for interpretation of the rule (\ref{b1.5}) and for
interpretation of the correspondence principle. The obtained results may be
applied to the quantum system $\mathcal{S}_{\mathrm{S}}$, because the
difference between the dynamic systems $\mathcal{S}_{\mathrm{S}}=\mathcal{E}%
\left[ \mathcal{S}_{\mathrm{st}}\right] $ and $\mathcal{S}_{\mathrm{Scl}}=%
\mathcal{E}\left[ \mathcal{S}_{\mathrm{d}}\right] $, described respectively
by actions (\ref{a1.2}) and (\ref{b1.21}), manifests itself only in the
additional nonlinear term in the dynamic equation. The possibility of
description $\mathcal{S}_{\mathrm{S}}=\mathcal{E}\left[ \mathcal{S}_{\mathrm{%
st}}\right] $ and $\mathcal{S}_{\mathrm{Scl}}=\mathcal{E}\left[ \mathcal{S}_{%
\mathrm{d}}\right] $ in both languages (classical and quantum) shuts the
door before the Copenhagen interpretation, where the wave function is
supposed to describe a single particle. Thus, \textit{there is neither
reason nor excuse for application of the Copenhagen interpretation}.

Mathematical formalism of quantum mechanics deals with continuous dynamic
systems: Schr\"{o}dinger particle $\mathcal{S}_{\mathrm{S}}$, Dirac particle 
$\mathcal{S}_{\mathrm{D}}$, etc. From the viewpoint of the mathematical
technique it is of no importance, whether $\mathcal{S}_{\mathrm{S}}$ is an
individual particle, or it is a statistical ensemble of particles
(statistically average particle). The Copenhagen interpretation meets the
difficulties, when it tries to test predictions of mathematical formalism in
single experiments. For instance, there exists the problem of the mechanism
of the wave function reduction in the single experiment. Another problem
concerns the two-slit experiment. How can an individual particle pass
through two slits at once? Such problems of quantum interpretations are not
wiredrawn. The physical journals publish discussions concerning problems of
quantum interpretation. For instance, such a discussion was declared in 2002
by the journal Uspekhi Fizicheskich Nauk. These problems cannot be solved in
the framework of the Copenhagen interpretation, which does not distinguish
between the individual particle $\mathcal{S}$ and the statistically average
particle $\left\langle \mathcal{S}\right\rangle $. Confusion of two
different objects, having different properties generates difficulties and
paradoxes. The wave function does not describe the state of individual
particle $\mathcal{S}$, and it is meaningless to ask, how the state of
individual particle changes at a single measurement (S-measurement). At the
massive experiment (M-measurement) we obtain a distribution $F\left(
R^{\prime }\right) $ of the measured quantity $\mathcal{R}$, but not a
single value $R\prime $ of the measured quantity. At such a situation it is
useless to ask, how the obtained result $R^{\prime }$ influences on the the
state of the statistical ensemble. Finally, we may define the third type of
measurement (SM-measurement): the massive measurement of the quantity $%
\mathcal{R}$ leading to a definite value $R^{\prime }$ of the measured
quantity $\mathcal{R}$. The SM-measurement is the M-measurement leading to a
distribution $F\left( R^{\prime }\right) $, accompanied by a selection of
those particles, where result of S-measurement is $R^{\prime }$. Uniting all
particles with the measured value $R^{\prime }$ in one statistical ensemble $%
\mathcal{E}_{R^{\prime }}$, we can put the question about the wave function
of $\mathcal{E}_{R^{\prime }}$. Of course, the wave function $\psi
_{R^{\prime }}$ of $\mathcal{E}_{R^{\prime }}$ does not coincide, in
general, with the initial wave function $\psi $, and this change of the wave
function is considered as a reduction of the wave function. The origin of
the reduction is quite transparent. It is the selection, which is produced
to obtain the same value $R^{\prime }$ of the measured quantity for all
particles of the statistical ensemble. Thus, the problems of reduction are
conditioned by the confusion of concepts of the individual particle $%
\mathcal{S}$ and the statistical average particle $\left\langle \mathcal{S}%
\right\rangle $, which takes place at the Copenhagen interpretion.

As concerns the particle, passing through two slits simultaneously, it is a
reasonable property of the statistical average object. It is a pure
statistical property, which has nothing to do with quantum properties.
Individual particle can pass either through one slit, or through another,
whereas the statistical average particle can pass through both slits
simultaneously. (Compare, individual person is either a man, or a woman,
whereas the statistical average person is a hermaphrodite (half-man
half-woman), and there are no quantum mechanical properties here.)

\section{Dynamic disquantization}

The quantum langauge, i.e. the description, containing the quantum constant $%
\hbar $, may be used for a description of a classical dynamic system,
because the quantum constant $\hbar $ may be used instead of the arbitrary
dynamical constant $b$. Replacement of dynamical constant $b$ by the quantum
constant is produced to make the dynamic equations to be linear. For
instance, in the action (\ref{a1.10}) the quantum constant $\hbar $ is used
naturally, i.e. in the sense that setting $\hbar =0$, we suppress the
quantum effects. In the action (\ref{a1.2}) for the same dynamic system the
quantum constant $\hbar $ is used artificially in the sense that setting $%
\hbar =0$, we do not suppress the quantum effects. Furthermore, setting $%
\hbar =0$, we destroy any description. But the action (\ref{a1.2}) generates
linear dynamic equation, and this circumstance is a reason for the
artificial identification $b=\hbar $, when the \textit{dynamical} constant $%
b $ is identified with the \textit{quantum} constant $\hbar $.

Such an artificial identification may be produced in other quantum systems
(for instance, in $\mathcal{S}_{\mathrm{D}}$ and $\mathcal{S}_{\mathrm{P}}$%
), and we cannot be sure, that setting $\hbar =0$, we suppress the quantum
effects. Besides, we cannot be sure that, using the transformation of the
type (\ref{a1.9}), we can separate the quantum terms from dynamical and
statistical ones.

We need a more effective formal dynamical procedure, which could suppress
the stochastic terms. Let us compare dynamic equations (\ref{b1.11}) for $%
\mathcal{S}_{\mathrm{S}}=\mathcal{E}\left[ \mathcal{S}_{\mathrm{st}}\right] $
written in the form 
\begin{equation}
\frac{d\mathbf{p}}{dt}=-\mathbf{\nabla }U\left( \rho ,\mathbf{\nabla }\rho
\right) ,\qquad \frac{d\mathbf{x}}{dt}=\frac{\mathbf{p}}{m},\qquad U\left(
\rho ,\mathbf{\nabla }\rho \right) =\frac{\hbar ^{2}}{8m}\left( \frac{\left( 
\mathbf{\nabla }\rho \right) ^{2}}{\rho ^{2}}-2\frac{\mathbf{\nabla }%
^{2}\rho }{\rho }\right)  \label{b2.2}
\end{equation}%
with the dynamic equations for $\mathcal{S}_{\mathrm{Scl}}=\mathcal{E}\left[ 
\mathcal{S}_{\mathrm{d}}\right] $, which have the form 
\begin{equation}
\frac{d\mathbf{p}}{dt}=0,\qquad \frac{d\mathbf{x}}{dt}=\frac{\mathbf{p}}{m}
\label{b2.1}
\end{equation}%
where $\mathbf{x}=\mathbf{x}\left( t,\mathbf{\xi }\right) $, $\mathbf{p}=%
\mathbf{p}\left( t,\mathbf{\xi }\right) $. Dynamic equations (\ref{b2.2}),
are the partial differential equations, because $\rho $ is defined by the
relation (\ref{b1.9}), containing derivatives with respect to $\xi _{\alpha
} $, $\alpha =1,2,3$, whereas dynamic equations (\ref{b2.1}) are ordinary
differential equations. Equations (\ref{b2.1}) contain derivatives only in
one direction in the space of independent variables $\left\{ t,\mathbf{\xi }%
\right\} $, whereas equations (\ref{b2.2}) contain derivatives in different
directions of the space of independent variables $\left\{ t,\mathbf{\xi }%
\right\} $. This property is conserved at any change of independent
dynamical variables, and, in particular, at the change $\left\{ t,\mathbf{%
\xi }\right\} \rightarrow \left\{ t,\mathbf{x}\right\} $. If a system of
partial differential equations contains derivative only in one direction of
the space of independent variables, this system can be reduced to the system
of ordinary differential equations by means of a proper change of variables.

Usually the classical particle (for instance, the classical Schr\"{o}dinger
particle) has two properties: (1) it does not contain the quantum constant $%
\hbar $, (2) it has the finite number of the freedom degrees. We may imagine
such a dynamic system, which contains the quantum constant $\hbar $ and has
definite number of the freedom degrees. Is such a dynamic system classical?
We believe that such a dynamic system should be classified as classical,
because it is deterministic dynamical system, but not a stochastic one. As
concerns to dependence on the quantum constant $\hbar $, it may describe, in
general, not only stochastic effects. The quantum constant $\hbar $ is an
attribute of the space-time \cite{R2005}, and in principle it may appear as
a parameter in deterministic classical systems, but not only in the
stochastic ones. Besides, the dynamic system $\mathcal{S}_{\mathrm{f}}$ with
finite number of the freedom degrees is simpler for investigation, than the
continuous dynamic system $\mathcal{S}_{\mathrm{c}}$, having infinite number
of the freedom degrees, because dynamic equations of $\mathcal{S}_{\mathrm{f}%
}$ are the ordinary differential equations, whereas dynamic equations of $%
\mathcal{S}_{\mathrm{c}}$ are partial differential equations. Transition to
the classical approximation as a means of investigation of the quantum
dynamic system, which is continuous, is justified only if the classical
system is the dynamic system $\mathcal{S}_{\mathrm{f}}$.

If we want to suppress the quantum effects, we must to reduce the system of
partial differential equations to the system of ordinary differential
equations. To make this, we should project derivatives in the space of
independent variables onto some direction. Then the system will contain
derivatives only in one direction, and hence it may be reduced to the system
of ordinary differential equations. Onto what direction should derivatives
in the system (\ref{b2.2}) be projected, to obtain the system (\ref{b2.1})?

Such a direction is described by the 4-current $j^{k}=\left\{ \rho ,\mathbf{j%
}\right\} =\left\{ j^{k}\right\} ,$ $k=0,1,2,3$ in the space-time. The
projection should be made in the space of independent variables $\left\{ t,%
\mathbf{x}\right\} $, i.e. in the space-time. It is convenient to choose
dependent variables in such a way, that the 4-current $j^{k}$ were one of
dependent variables. We take the action (\ref{A.12}) for the dynamic system $%
\mathcal{S}_{\mathrm{S}}=\mathcal{E}\left[ \mathcal{S}_{\mathrm{st}}\right] $%
\begin{equation}
\mathcal{A}_{\mathcal{E}\left[ \mathcal{S}_{\mathrm{st}}\right] }\left[
\varphi ,\mathbf{\xi },j\right] =\int \left\{ \frac{m}{2}\frac{j^{\alpha
}j^{\alpha }}{\rho }-bj^{k}\left( \partial _{k}\varphi +g^{\alpha }\left( 
\mathbf{\xi }\right) \partial _{k}\xi _{\alpha }\right) -\frac{\hbar ^{2}}{8m%
}\frac{\left( \mathbf{\nabla }\rho \right) ^{2}}{\rho }\right\} d^{4}x%
\mathbf{,}  \label{b2.3}
\end{equation}%
where according to (\ref{A.17}) and (\ref{g1.30}) 
\begin{equation}
j^{k}=\left\{ \rho ,\mathbf{j}\right\} =\left\{ \rho ,\frac{b\rho }{m}\left( 
\mathbf{\nabla }\varphi +g^{\alpha }\left( \mathbf{\xi }\right) \mathbf{%
\nabla }\xi _{\alpha }\right) \right\}  \label{b2.4}
\end{equation}%
and $g^{\alpha }\left( \mathbf{\xi }\right) $, $\alpha =1,2,3$ are arbitrary
functions of argument $\mathbf{\xi }$.

The second term in the action (\ref{b2.3}) contains derivatives only in the
direction of the 4-vector $j^{k}$. In the last term of (\ref{b2.3}) the
derivatives are to be projected onto the vector $j^{k}$. We are to make the
change 
\begin{equation}
\partial _{l}\rightarrow \partial _{||l}=\frac{j_{l}j^{k}}{j_{s}j^{s}}%
\partial _{k},\qquad l=0,1,2,3  \label{b2.5}
\end{equation}
in the action (\ref{b2.3}). We obtain 
\begin{equation}
\frac{\left( \mathbf{\nabla }\rho \right) ^{2}}{\rho }=\frac{\left( \mathbf{%
\partial }_{\alpha }\rho \right) \left( \mathbf{\partial }_{\alpha }\rho
\right) }{\rho }\rightarrow \frac{j_{\alpha }j_{\alpha }\left( j^{i}\partial
_{i}\rho \right) ^{2}}{\rho \left( j^{s}j_{s}\right) ^{2}}  \label{b2.6}
\end{equation}
\begin{equation*}
j_{\alpha }j_{\alpha }=\mathbf{j}^{2}=\rho ^{2}\mathbf{v}^{2},\qquad
j^{s}j_{s}=c^{2}\rho ^{2}-\rho ^{2}\mathbf{v}^{2}
\end{equation*}
In the nonrelativistic approximation, when the velocity $\left\vert \mathbf{v%
}\right\vert \ll c$, we obtain the following estimation 
\begin{equation}
\frac{\left( \mathbf{\nabla }\rho \right) ^{2}}{\rho }\approx \frac{\mathbf{v%
}^{2}\left( j^{i}\partial _{i}\rho \right) ^{2}}{c^{4}\rho ^{3}}
\label{b2.7}
\end{equation}
In the nonrelativistic approximation $c\rightarrow \infty $ the last term in
the action (\ref{b2.3}) is to be neglected after the change (\ref{b2.5}).
Thus, in the case of the Schr\"{o}dinger particle $\mathcal{S}_{\mathrm{S}}$
the change (\ref{b2.5}) leads to a suppression of quantum effects.

We shall refer to the procedure (\ref{b2.5}) as the dynamical
disquantization, because it transforms the Schr\"{o}dinger particle $%
\mathcal{S}_{\mathrm{S}}=\mathcal{E}\left[ \mathcal{S}_{\mathrm{st}}\right] $
into the classical system $\mathcal{S}_{\mathrm{Scl}}=\mathcal{E}\left[ 
\mathcal{S}_{\mathrm{d}}\right] $. The dynamical disquantization is the
relativistic dynamical procedure. It does not refer to the quantum constant
and suppresses any stochasticity regardless of its origin. From here on we
shall use the dynamical disquantization for the suppression of stochasticity
in quantum systems.

Strictly, the dynamical disquantization is to be applied to the dynamic
equations. But in many cases the application of the dynamical
disquantization to the action leads to the same result, as its application
to the dynamic equations.

\section{Classical approximation of the Dirac particle}

For transition to the classical approximation it is sufficient to make
dynamical disquantization of the Dirac equation (\ref{f1.2}). Using
expression (\ref{f1.3}) for 4-current $j^{k}$, we make this directly in (\ref%
{f1.2}). We obtain%
\begin{equation}
-i\hbar \frac{\psi ^{\ast }\gamma ^{0}\gamma _{l}\psi }{\left( \psi ^{\ast
}\gamma ^{0}\gamma ^{s}\psi \right) \left( \psi ^{\ast }\gamma ^{0}\gamma
_{s}\psi \right) }\left( \psi ^{\ast }\gamma ^{0}\gamma ^{k}\psi \right)
\gamma ^{l}\partial _{k}\psi +\frac{e}{c}A_{l}\gamma ^{l}\psi +mc\psi =0
\label{a4.1}
\end{equation}%
or after transformation%
\begin{equation}
-i\hbar \frac{d}{d\tau }\psi +\frac{e}{c}A_{l}\frac{j_{k}}{\sqrt{j^{s}j_{s}}}%
\gamma ^{k}\gamma ^{l}\psi +mc\frac{j_{k}}{\sqrt{j^{s}j_{s}}}\gamma ^{k}\psi
=0,\qquad \frac{d}{d\tau }\equiv \frac{j^{k}}{\sqrt{j^{s}j_{s}}}\partial _{k}
\label{a4.2}
\end{equation}%
where $j^{k}$ is determined by the relation (\ref{f1.3}) via the wave
function $\psi $.

Formally equations (\ref{a4.2}) form a system of ordinary differential
equations for the dependent variables $\psi $, considered as function of
independent variable $\tau $. Unfortunately, interpretation of equations (%
\ref{a4.2}) is difficult. It is more convenient to use the hydrodinamical
variables, where the the components of the 4-current $j^{k}$ are dependent
variables. The transformation to the hydrodynamical variables may be easier
carried out in the action (\ref{b1.1}).

Transforming the action (\ref{b1.1}), we use the mathematical technique \cite%
{S30,S51}, where the wave function $\psi $ is considered to be a function of
hypercomplex numbers $\gamma $ and coordinates $x$. In this case the
physical quantities are obtained by means of a convolution of expressions $%
\psi ^{\ast }O\psi $ with the zero divisor. This technique allows one to
work without fixing the $\gamma $-matrices representation.

Using designations 
\begin{equation}
\gamma _{5}=c\gamma ^{0123}\equiv c\gamma ^{0}\gamma ^{1}\gamma ^{2}\gamma
^{3},  \label{f1.9}
\end{equation}%
\begin{equation}
\mathbf{\sigma }=\{\sigma _{1},\sigma _{2},\sigma _{3},\}=\{-i\gamma
^{2}\gamma ^{3},-i\gamma ^{3}\gamma ^{1},-i\gamma ^{1}\gamma ^{2}\}
\label{f1.10}
\end{equation}%
we make the change of variables 
\begin{equation}
\psi =Ae^{i\varphi +{\frac{1}{2}}\gamma _{5}\kappa }\exp \left( -\frac{i}{2}%
\gamma _{5}\mathbf{\sigma \eta }\right) \exp \left( {\frac{i\pi }{2}}\mathbf{%
\sigma n}\right) \Pi  \label{f1.11}
\end{equation}%
\begin{equation}
\psi ^{\ast }=A\Pi \exp \left( -{\frac{i\pi }{2}}\mathbf{\sigma n}\right)
\exp \left( -\frac{i}{2}\gamma _{5}\mathbf{\sigma \eta }\right) e^{-i\varphi
-{\frac{1}{2}}\gamma _{5}\kappa }  \label{f1.12}
\end{equation}%
where (*) means the Hermitian conjugation, and the quantity 
\begin{equation}
\Pi ={\frac{1}{4}}(1+c\gamma ^{0})(1+\mathbf{z\sigma }),\qquad \mathbf{z}%
=\{z^{\alpha }\}=\text{const},\qquad i\alpha =1,2,3;\qquad \mathbf{z}^{2}=1
\label{f1.13}
\end{equation}%
is the zero divisor (projector). The quantities $A$, $\kappa $, $\varphi $, $%
\mathbf{\eta }=\{\eta ^{\alpha }\}$, $\mathbf{n}=\{n^{\alpha }\}$, $\alpha
=1,2,3,\;$ $\mathbf{n}^{2}=1$ are eight real parameters, determining the
wave function $\psi .$ These parameters may be considered as new dependent
variables, describing the state of dynamic system $\mathcal{S}_{\mathrm{D}}$%
. The quantity $\varphi $ is a scalar, and $\kappa $ is a pseudoscalar. Six
remaining variables $A,$ $\mathbf{\eta }=\{\eta ^{\alpha }\}$, $\mathbf{n}%
=\{n^{\alpha }\}$, $\alpha =1,2,3,\;$ $\mathbf{n}^{2}=1$ can be expressed
through the flux 4-vector $j^{l}=\bar{\psi}\gamma ^{l}\psi $ and spin
4-pseudovector 
\begin{equation}
S^{l}=i\bar{\psi}\gamma _{5}\gamma ^{l}\psi ,\qquad l=0,1,2,3  \label{f1.13a}
\end{equation}%
Because of two identities 
\begin{equation}
S^{l}S_{l}\equiv -j^{l}j_{l},\qquad j^{l}S_{l}\equiv 0.  \label{f1.14}
\end{equation}%
there are only six independent components among eight components of
quantities $j^{l}$, and $S^{l}$.

Mathematical details of the dependent variables transformation can be found
in \cite{R2004}, where the action is calculated for the case $c=1$ and
vanishing electromagnetic field $A_{l}=0$. As a result we have the following
form of the action, written in the hydrodynamical form 
\begin{equation}
\mathcal{S}_{\mathrm{D}}:\qquad \mathcal{A}_{D}[j,\varphi ,\kappa ,\mathbf{%
\xi }]=\int \mathcal{L}d^{4}x,\qquad \mathcal{L}=\mathcal{L}_{\mathrm{cl}}+%
\mathcal{L}_{\mathrm{q1}}+\mathcal{L}_{\mathrm{q2}}  \label{c4.15}
\end{equation}%
\begin{equation}
\mathcal{L}_{\mathrm{cl}}=-mc\rho -\hbar j^{l}\partial _{l}\varphi -\frac{e}{%
c}A_{l}j^{l}-\frac{\hbar j^{l}}{2\left( 1+\mathbf{\xi z}\right) }\varepsilon
_{\alpha \beta \gamma }\xi ^{\alpha }\partial _{l}\xi ^{\beta }z^{\gamma
},\qquad \rho \equiv \sqrt{j^{l}j_{l}}  \label{c4.16}
\end{equation}%
\begin{equation}
\mathcal{L}_{\mathrm{q1}}=2mc\rho \sin ^{2}(\frac{\kappa }{2})-{\frac{\hbar 
}{2}}S^{l}\partial _{l}\kappa ,  \label{c4.17}
\end{equation}%
\begin{equation}
\mathcal{L}_{\mathrm{q2}}=\frac{\hbar (\rho +cj^{0})}{2}\varepsilon _{\alpha
\beta \gamma }\partial ^{\alpha }\frac{j^{\beta }}{(\rho +cj^{0})}\xi
^{\gamma }-\frac{\hbar }{2(\rho +cj^{0})}\varepsilon _{\alpha \beta \gamma
}\left( \partial ^{0}j^{\beta }\right) j^{\alpha }\xi ^{\gamma }
\label{c4.18}
\end{equation}%
where $\varepsilon _{\alpha \beta \gamma }$ is the Levi-Chivita
3-pseudotensor. The Lagrangian density $\mathcal{L}$ is a function of
4-vector $j^{l}$, scalar $\varphi $, pseudoscalar $\kappa $, and the unit
3-pseudovector $\mathbf{\xi }$, which is connected with the spin
4-pseudovector $S^{l}$ by means of the relations 
\begin{equation}
\xi ^{\alpha }=\rho ^{-1}\left[ S^{\alpha }-\frac{j^{\alpha }S^{0}}{(\rho
+cj^{0})}\right] ,\qquad \alpha =1,2,3;\qquad \rho \equiv \sqrt{j^{l}j_{l}}
\label{f1.15}
\end{equation}%
\begin{equation}
S^{0}=\mathbf{j\xi },\qquad S^{\alpha }=\rho \xi ^{\alpha }+\frac{(\mathbf{%
j\xi })j^{\alpha }}{\rho +cj^{0}},\qquad \alpha =1,2,3  \label{f1.16}
\end{equation}

Let us set for simplicity $A_{l}=0$ and $c=1$. Producing the dynamical
disquantization (\ref{b2.5}) in (\ref{c4.15}) - (\ref{c4.18}), we obtain%
\begin{equation*}
\mathcal{A}_{\mathrm{Dqu}}[j,\varphi ,\mathbf{\xi }]=\int \left\{ -\kappa
_{0}m\rho -\hbar j^{i}\left( \partial _{i}\varphi +\frac{\varepsilon
_{\alpha \beta \gamma }\xi _{\alpha }\partial _{i}\xi _{\beta }z_{\gamma }}{%
2\left( 1+\mathbf{\xi z}\right) }\right) \right.
\end{equation*}
\begin{equation}
+\left. \frac{\hbar j^{k}}{2(\rho +j_{0})\rho }\varepsilon _{\alpha \beta
\gamma }\left( \partial _{k}j^{\beta }\right) j^{\alpha }\xi _{\gamma
}\right\} d^{4}x  \label{a5.13a}
\end{equation}%
where $\kappa _{0}=\pm 1$ is the solution of the dynamic equation $\delta 
\mathcal{A}_{\mathrm{Dqu}}/\delta \kappa =0$, which does not contain
derivatives, because the last term of (\ref{c4.17}) vanishes after dynamical
disquantization (\ref{b2.5}) in virtue of the second identity (\ref{f1.14}).

We introduce the Lagrangian coordinates $\tau =\{\tau _{0},\mathbf{\tau }%
\}=\{\tau _{i}\left( x\right) \}$, $i=0,1,2,3$ as functions of coordinates $%
x $ in such a way that only coordinate $\tau _{0}$ changes along the
direction $j^{l}$, i.e. 
\begin{equation}
j^{k}\partial _{k}\tau _{\mu }=0,\qquad \mu =1,2,3  \label{b3.1}
\end{equation}%
Considering the variables $\tau =\{\tau _{0},\mathbf{\tau }\}$ as
independent variables in (\ref{a5.13a}), we obtain after calculations (See
mathematical details in \cite{R2004}) 
\begin{equation}
\mathcal{A}_{\mathrm{Dqu}}[x,\mathbf{\xi }]=\int \left\{ -\kappa _{0}m\sqrt{%
\dot{x}^{l}\dot{x}_{l}}+\hbar {\frac{(\dot{\mathbf{\xi }}\times \mathbf{\xi }%
)\mathbf{z}}{2(1+\mathbf{\xi z})}}+\hbar \frac{(\dot{\mathbf{x}}\times \ddot{%
\mathbf{x}})\mathbf{\xi }}{2\sqrt{\dot{x}^{s}\dot{x}_{s}}(\sqrt{\dot{x}^{s}%
\dot{x}_{s}}+\dot{x}^{0})}\right\} d^{4}\tau  \label{a5.18}
\end{equation}%
where period means the total derivative $\dot{x}^{s}\equiv dx^{s}/d\tau _{0}$%
. The quantities \ $x=\left\{ x^{0},\mathbf{x}\right\} =\{x^{i}\}$, $%
\;i=0,1,2,3$, and $\mathbf{\xi }=\{\xi _{\alpha }\}$, $\alpha =1,2,3$ are
considered to be functions of the Lagrangian coordinates $\tau _{0}$, $%
\mathbf{\tau }=\{\tau _{1},\tau _{2},\tau _{3}\}$. Here and in what follows
the symbol $\times $ means the vector product of two 3-vectors. The quantity$%
\;\mathbf{z}$ is the constant unit 3-vector (\ref{f1.13}). In fact,
variables $x$ depend on $\mathbf{\tau }$ as on parameters, because the
action (\ref{a5.18}) does not contain derivatives with respect to $\tau
_{\alpha }$, \ $\alpha =1,2,3$. Lagrangian density of the action (\ref{a5.18}%
) does not contain independent variables $\tau $ explicitly. Hence, it may
be written in the form 
\begin{equation}
\mathcal{A}_{\mathrm{Dqu}}[x,\mathbf{\xi }]=\int \mathcal{A}_{\mathrm{Dcl}%
}[x,\mathbf{\xi }]d\mathbf{\tau ,\qquad }d\mathbf{\tau }=d\tau _{1}d\tau
_{2}d\tau _{3}  \label{b3.8}
\end{equation}%
where 
\begin{equation}
\mathcal{A}_{\mathrm{Dcl}}[x,\mathbf{\xi }]=\int \left\{ -\kappa _{0}mc\sqrt{%
\dot{x}^{i}\dot{x}_{i}}+\hbar {\frac{(\dot{\mathbf{\xi }}\times \mathbf{\xi }%
)\mathbf{z}}{2(1+\mathbf{\xi z})}}+\hbar \frac{(\dot{\mathbf{x}}\times \ddot{%
\mathbf{x}})\mathbf{\xi }}{2\sqrt{\dot{x}^{s}\dot{x}_{s}}(\sqrt{\dot{x}^{s}%
\dot{x}_{s}}+\dot{x}^{0})}\right\} d\tau _{0}  \label{b3.9}
\end{equation}%
Here period means differentiation with respect to $\tau _{0}$. It is easy to
see that the action (\ref{b3.9}) is invariant with respect to transformation 
$\tau _{0}\rightarrow \tilde{\tau}_{0}=F\left( \tau _{0}\right) $, where $F$
is an arbitrary monotone function.

The action (\ref{b3.9}) can be written in the relativistically covariant
form 
\begin{equation}
\mathrm{\mathcal{A}}_{\mathrm{Dcl}}\left[ x,\xi \right] =\int \{-\kappa _{0}m%
\sqrt{\dot{x}^{i}\dot{x}_{i}}-\frac{e}{c}A_{l}\dot{x}^{l}-\hbar \frac{%
\varepsilon _{iklm}\xi ^{i}\dot{\xi}^{k}f^{l}z^{m}}{2(1-\xi ^{s}z_{s})}+%
\frac{\hbar }{2}Q\varepsilon _{iklm}\dot{x}^{i}\ddot{x}^{k}f^{l}\xi
^{m}\}d\tau _{0}  \label{f5.1}
\end{equation}%
\begin{equation}
Q=Q\left( \dot{x},f\right) =\frac{1}{\sqrt{\dot{x}^{s}\dot{x}_{s}}(\dot{x}%
^{l}f_{l}+\sqrt{\dot{x}^{l}\dot{x}_{l}})},\qquad \dot{x}^{l}\equiv \frac{%
dx^{l}}{d\tau _{0}}  \label{f5.2}
\end{equation}%
where 4-vectors $f^{k}$, $z^{k}$ are defined respectively by relations 
\begin{equation}
f^{i}=\{1,0,0,0\}  \label{d2.1}
\end{equation}%
\begin{equation}
z^{k}=\left\{ 0,\mathbf{z}\right\} =\left( 0,z^{1},z^{2},z^{3}\right)
,\qquad z^{k}z_{k}=-1,\qquad \mathbf{\xi z}=-\xi _{l}z^{l}  \label{d2.8}
\end{equation}%
\begin{equation}
\xi ^{k}=\left\{ \xi _{0},\mathbf{\xi }\right\} ,\qquad \xi ^{l}f_{l}=0
\label{f5.2a}
\end{equation}

The quantities $f^{k}$ and $z^{k}$ are constant 4-vectors. They have the
form (\ref{d2.1}) and (\ref{d2.8}) only in some coordinate system. In other
coordinate systems their form is obtained as a result of corresponding
transformation of relations (\ref{d2.1}) and (\ref{d2.8}). These 4-vectors
describe some structures, existing in the space-time. The vector $z^{k}$
describes introduction of the projector (\ref{f1.13}). The 4-vector $z^{k}$
appears to be fictitious (see \cite{R2004}).

However, the 4-vector $f^{k}$ is not fictitious. It describes existence of
some special direction in the space-time. This direction describes
separation of the space-time into absolute time and absolute space. Having
the vector $f^{k}$, one can assign absolute time $t=x^{k}f_{k}$ and absolute
spatial distance $r=\sqrt{\left( x^{k}f_{k}\right) ^{2}-x^{k}x_{k}}$ to any
space-time vector $x^{k}$. It means that the terms, containing the constant
4-vector $f^{k}$ are nonrelativistic. This statement is in accordance with
the theorem, proved by Anderson \cite{A67}. According to this theorem the
symmetry group of dynamic equations, written in the relativistically
covariant form is determined by the symmetry group of absolute objects. The
absolute objects are the quantities, which are the same for all solutions of
the dynamic equations. The absolute objects are structures or force fields
external with respect to considered dynamic system. In the given case the
4-vector $f^{k}$ is such an absolute object. If we set (\ref{d2.1}) in (\ref%
{f5.1}), we return to the action (\ref{a5.18}), which does not contain the
absolute object $f^{k}$, but simultaneously the action (\ref{a5.18}) appears
to be written in the noncovariant form. In other words, the relativistically
covariant form of description is a method of separation of absolute objects
(or space-time structures) used at the description of the dynamic system.

Two first terms in the action (\ref{f5.1}) do not contain the 4-vector $%
f^{k} $. They describe relativistically the motion of the classical Dirac
particle as a whole. Two last terms contain the 4-vector $f^{k}$ and the
4-pseudovector $\xi ^{k}$. They describe internal degrees of freedom of the
classical Dirac particle. The description of the internal degrees of freedom
is nonrelativistic.

Dynamical methods of investigation and the formalized procedure of
transition to the classical approximation (\ref{b2.5}) admit one to discover
two unexpected properties of the Dirac particle: (1) the Dirac particle has
an internal degrees of freedom, (2) these degrees of freedom are described
nonrelativistically. The last property means that formally the Dirac
equation is nonrelativistic equation, i.e. the set of all solutions of the
Dirac equation is not invariant with respect to the Lorentz group of
transformations.

In the case, when the electromagnetic field is absent $A_{k}=0$, dynamic
equations, generated by the action (\ref{b3.9}) have the form (see details
in \cite{R2004})%
\begin{equation}
\frac{d}{d\tau _{0}}\left( -\kappa _{0}m\frac{\mathbf{\dot{x}}}{\sqrt{\dot{x}%
^{s}\dot{x}_{s}}}+\frac{\hbar Q}{2}(\mathbf{\xi }\times \ddot{\mathbf{x}})-%
\frac{\hbar }{2}\frac{\partial Q}{\partial \mathbf{\dot{x}}}(\dot{\mathbf{x}}%
\times \ddot{\mathbf{x}})\mathbf{\xi }+\frac{\hbar }{2}\frac{d}{d\tau _{0}}%
\left( Q(\mathbf{\xi }\times \mathbf{\dot{x}})\right) \right) =0
\label{b6.1}
\end{equation}%
\begin{equation}
\frac{d}{d\tau _{0}}\left( \kappa _{0}m\frac{\dot{x}^{0}}{\sqrt{\dot{x}^{s}%
\dot{x}_{s}}}-\frac{\hbar }{2}\frac{\partial Q}{\partial \dot{x}^{0}}(\dot{%
\mathbf{x}}\times \ddot{\mathbf{x}})\mathbf{\xi }\right) =0  \label{b6.1a}
\end{equation}%
\begin{equation}
\mathbf{\dot{\xi}}=-(\mathbf{\dot{x}}\times \mathbf{\ddot{x}})\times \mathbf{%
\xi }Q,  \label{c7.1}
\end{equation}%
where the quantity $Q$ is defined by the relation (\ref{f5.2}), (\ref{d2.1}%
), i.e. 
\begin{equation}
Q=Q\left( \dot{x}\right) =\left( \sqrt{\dot{x}^{s}\dot{x}_{s}}(\sqrt{\dot{x}%
^{s}\dot{x}_{s}}+\dot{x}^{0})\right) ^{-1},\qquad \dot{x}^{s}\dot{x}_{s}=%
\dot{x}_{0}^{2}-\mathbf{\dot{x}}^{2}  \label{b6.2}
\end{equation}%
World line of the classical Dirac particle, described by the dynamic
equations (\ref{b6.1}), (\ref{b6.1a}), (\ref{c7.1}), is a helix (see details
in \cite{R2004})%
\begin{equation}
\frac{d\mathbf{x}}{dt}=\left\{ \frac{\sqrt{\gamma ^{2}-1}}{\gamma }\cos
\left( \Omega t\right) ,-\frac{\sqrt{\gamma ^{2}-1}}{\gamma }\sin \left(
\Omega t\right) ,0\right\} ,\qquad \Omega =\frac{2m}{\hbar \gamma ^{2}}
\label{e6.30}
\end{equation}%
\begin{equation}
\mathbf{x}=\left\{ \frac{\hbar \gamma \sqrt{\gamma ^{2}-1}}{2m}\sin \left( 
\frac{2m}{\hbar \gamma ^{2}}t\right) ,\frac{\hbar \gamma \sqrt{\gamma ^{2}-1}%
}{2m}\cos \left( \frac{2m}{\hbar \gamma ^{2}}t\right) ,0\right\}
\label{e6.31}
\end{equation}%
where $\gamma \geq 1$ is an arbitrary constant (Lorentz factor), describing
intensity of the circular motion of the classical Dirac particle.

Details of interpretation of solution (\ref{e6.31}) can be found in \cite%
{R2004}. We shall not go into these details here, because we are interested
mainly in interplay between the dynamical methods and conventional axiomatic
methods of the quantum system investigation. In particular, we are
interested in such questions. Is it possible to discover internal degrees of
freedom of the Dirac particle by conventional methods? Can one discover the
nonrelativistical character of the Dirac equation by conventional methods?
If yes, then why have not these properties of the Dirac particle been
discovered for eighty years? In other words, we want to compare
effectiveness of dynamical methods and axiomatic ones. There was a reason
for the statement of such questions.

\section{On mistakes in application of the conventional investigation method
to the Dirac particle}

At first, the unexpected properties of the Dirac particle were presented in
the paper \cite{R2001}. Unfortunately, I failed to publish this paper in a
physical journal, because of negative review of referees. Statements of the
paper on internal degrees of freedom and especially on a nonrelativistic
character of the Dirac equation met objections of the referees. These
objections were of such a kind: (1) 'It is well known that the Dirac
equation is relativistic', (2) 'Maybe, the author's calculations contain a
mistake, because the Dirac equation agrees very well with the experiments',
(3) 'I cannot imagine, that the Dirac equation is nonrelativistic, because
it can be written in the relativistically covariant form'. Discussion and
evaluation of the merits of the paper was absent in all reviews. The
viewpoint of referees reflects the statistical average opinion of the
scientific community, and it should be taken into account. The paper was
rejected despite my comments to the reviews of the referees.

Then I decided that two new unexpected results were too many for one paper
and divided the paper into two parts. In the paper \cite{R2004} the internal
degrees of freedom were discussed, whereas the paper \cite{R2004a} was
devoted to discussion of the nonrelativistic character of the Dirac
equation. Again the referees have not found mistakes or defects in my
papers, but they recommended against publication of the papers.

If the referees do not want or cannot find mistakes in my papers, I have to
search for mistakes in the conventional axiomatic method, which lead to the
absence of internal degrees of freedom of the Dirac particle and to the
relativistic character of the Dirac equation. The corresponding mistakes
have been easily found. Discovery and discussion of these mistakes is
presented below.

First about internal degrees of freedom of the Dirac particle.
Conventionally the Pauli equation is considered to be a nonrelativistic
approximation of the Dirac equation. The Pauli equation is a system of two
first order complex differential equations for two complex dependent
variables, whereas the Dirac equation is a system of four first order
complex differential equations for four complex dependent variables. It
means that in the classical approximation the classical Pauli particle $%
\mathcal{S}_{\mathrm{Pcl}}$ has less degrees of freedom, than the classical
Dirac particle $\mathcal{S}_{\mathrm{Dcl}}$. The Pauli particle $\mathcal{S}%
_{\mathrm{Pcl}}$ is the nonrelativistic pointlike particle with spin. Is the
classical Dirac particle $\mathcal{S}_{\mathrm{Dcl}}$ a relativistic
particle with spin and additional degrees of freedom, or is the order of the
Dirac system reduced indeed at the transition to the nonrelativistic
approximation?

Why is the order of the system of dynamic equations reduced at the
nonrelativistic approximation? It is reduced because coefficients before
some time derivatives are of the order $c^{-2}$, which vanish at $%
c\rightarrow \infty $. In other words, the Dirac system of dynamic equations
is a system of differential equations with small parameter before the
highest derivative. One may not neglect these terms, because at high
temporal frequency they may be of the same order as other terms, even if
coefficients before the derivatives are small. Neglecting the terms with
highest derivatives, we loss the high frequency solutions of dynamic
equations. From formal viewpoint such a neglect of the terms with highest
derivatives is a mathematical mistake, leading to a loss of the high
frequency solutions.

The scrupulous analysis shows \cite{R2005a} that solution of the Dirac
equation in the nonrelativistic approximation has the form%
\begin{equation}
\psi =\frac{\exp \left( -\frac{i}{2}\Omega t\right) }{\sqrt{2}}\left( 
\begin{array}{l}
\psi _{1}+\psi _{2} \\ 
\psi _{1}-\psi _{2}%
\end{array}%
\right) ,\qquad \Omega =\frac{2mc^{2}}{\hbar }  \label{B.3}
\end{equation}%
where $\psi $ is the dependent variable of the action (\ref{b1.1}), $\psi
_{1}$ and $\psi _{2}$ are two-component complex quantities, defined by the
relations%
\begin{equation}
\psi _{1}=u_{+}\left( t,\mathbf{x}\right) +e^{i\Omega t}u_{-}\left( t,%
\mathbf{x}\right)  \label{B.20}
\end{equation}%
\begin{equation}
\psi _{2}=-\frac{1}{2mc}\left( i\hbar \partial _{\mu }-\frac{e}{c}A_{\mu
}\right) \sigma _{\mu }\left( u_{+}\left( t,\mathbf{x}\right) +e^{i\Omega
t}u_{-}\left( t,\mathbf{x}\right) \right)  \label{B.18a}
\end{equation}%
Here $\sigma _{\mu }$, $\mu =1,2,3$ are the Pauli matrices, and $u_{+}\left(
t,\mathbf{x}\right) $, $u_{-}\left( t,\mathbf{x}\right) $ are two-component
quantities, which are solutions of the Pauli equations with the different
sign of the mass $m$.%
\begin{equation}
i\hbar \partial _{0}u_{+}=\hat{H}_{\mathrm{P}}\left( m\right) u_{+},\qquad
i\hbar \partial _{0}u_{-}=\hat{H}_{\mathrm{P}}\left( -m\right) u_{-}
\label{B.17a}
\end{equation}%
\begin{equation}
\hat{H}_{\mathrm{P}}\left( m\right) =\frac{\pi _{\mu }\pi _{\mu }}{2m}+\frac{%
ie\hbar }{2mc}\varepsilon _{\nu \mu \alpha }\partial _{\nu }A_{\mu }\sigma
_{\alpha }=\frac{\mathbf{\pi }^{2}}{2m}+\frac{ie\hbar }{2mc}\mathbf{H\sigma }
\label{B.8}
\end{equation}%
\begin{equation*}
\mathbf{\pi }=\left\{ \pi _{1},\pi _{2},\pi _{3}\right\} ,\qquad \pi _{\mu
}\equiv i\hbar \partial _{\mu }+\frac{e}{c}A_{\mu },\qquad \mu =1,2,3
\end{equation*}%
Here $\mathbf{H}=\mathbf{\nabla }\times \mathbf{A}$ is the magnetic field.

The frequency $\Omega =mc^{2}/\hbar $ tends to $\infty $, if $c\rightarrow
\infty $. In the nonrelativistic case, when $\gamma $ is close to $1$, the
frequency $\Omega $ coincides with the frequency $\Omega $, which figures in
relations (\ref{e6.30}), describing world line of the classical Dirac
particle $\mathcal{S}_{\mathrm{Dcl}}$. If we ignore high frequency part of
solution, setting $u_{-}=0$ in relations (\ref{B.3}) - (\ref{B.18a}), we
obtain a solution of the Pauli equation. As far as the Dirac particle is a
charged particle, the states, where $u_{+}\neq 0\wedge u_{-}\neq 0$, are
unstable, because at such a state the 4-current $j^{k}$ has components,
oscillating with the frequency $\Omega $. The Dirac particle emanates
electromagnetic waves of the frequency $\Omega $, until one of quantities $%
u_{+}$ or $u_{-}$ becomes to be equal to zero \cite{R2005a}. As far as the
time of transition to the stable state is very short, the Dirac particle
exists practically either at the low frequency state $\psi _{\mathrm{lf}}$,
when $u_{-}=0$, or at the high frequency state $\psi _{\mathrm{hf}}$, when $%
u_{+}=0$. The state $\psi _{\mathrm{lf}}$ is associated with the particle,
whereas the state $\psi _{\mathrm{hf}}$ is associated with the antiparticle.
Formally the superposition of the states $\psi _{\mathrm{lf}}$ and $\psi _{%
\mathrm{hf}}$ is possible, but it is not realized for the charged Dirac
particle because of its instability. This conclusion on the instability is
valid also for the classical Dirac particle $\mathcal{S}_{\mathrm{Dcl}}$,
described by the relations (\ref{e6.30}), (\ref{e6.31}).

Thus, the Dirac particle $\mathcal{S}_{\mathrm{D}}$, described by the action
(\ref{b1.1}), has internal degrees of freedom, which has not been discovered
because of a mathematical mistake in the transition to the nonrelativistic
approximation, which carries out an interpretation of the Dirac particle.
The internal degrees of freedom are not displayed in the nonrelativistic
applications of the Dirac equation, and there was no necessity of looking
for them. The transition to classical approximation must discover all
degrees of freedom automatically, but only in the case, when the procedure
of transition to the classical approximation is formalized. Conventionally
the transition to the classical approximation was not formalized, and there
are reason for this. It is accompanied by a series of additional
suppositions, which always can be chosen in such a way to obtain the
desirable result known in advance. First the procedure of transition to the
classical approximation was formalized in \cite{R2001}, and the internal
degrees of freedom of $\mathcal{S}_{\mathrm{Dcl}}$ were discovered
immidiately.

The Dirac equation can be written in the relativistically covariant form. It
is common practice to think, that it means that the Dirac equation a
relativistic equation. In other words, it and has the Lorentz symmetry, i.e.
the set of all its solutions is transformed to the same set of solutions at
any Lorentz transformation. This opinion has been existing for many years,
and we try to understand the reason of this viewpoint.

Formally the relativistic character of dynamic equations associates with the
representation of these equations in the relativistically covariant form.
However, this association is valid only at some additional conditions, which
are fulfilled practically always, and as a result these conditions are not
mentioned usually in the conventional formulation of the relativistic
invariance (i.e. compatibility of dynamic equations with the principles of
relativity). Unfortunately, in the case of the Dirac equation these
additional conditions are not fulfilled, and the Dirac equation appears to
be formally nonrelativistic. In reality, only internal degrees of freedom
are nonrelativistic. If these internal degrees of freedom are ignored, the
Dirac particle appears to be relativistic.

The additional constraint in the formulation of the relativistical
invariance changes the formulation. The correct formulation looks as
follows. \textit{Symmetry of dynamic equations, written in the
relativistically covariant form coincides with the symmetry of their
absolute objects} \cite{A67}. The absolute objects are such quantities,
which are the same for all solutions. Formally absolute objects are some
structures or objects, which are external with respect to considered dynamic
system. Usually such an absolute object is the metric tensor, which has the
form $g_{ik}=$diag$\left\{ c^{2},-1,-1,-1\right\} $. The group of symmetry
of $g_{ik}$ is the Lorentz group, and the symmetry group of dynamic
equations appears to be the Lorentz group. The Maxwell equations, the
Klein-Gordon equation and many other dynamic equations for real dynamic
systems contain only the metric tensor as an absolute object, and the
formulation of relativistical invariance is simplified. It looks as follows. 
\textit{The Lorentz group is the symmetry group of dynamic equations,
written in the relativistically covariant form}. In such a form it used by
most researchers.

The Dirac equation does not contain the metric tensor. Instead it contains
the $\gamma $-matrices $\gamma ^{i}$, $i=0,1,2,3$. The $\gamma $-matrices
form a matrix 4-vector, whose symmetry group is lower, than the Lorentz
group. As a result the Dirac equation appears to have not a symmetry of the
Lorentz group. In other words, the Dirac equation appears to be
nonrelativistic equation.

What physical situation is behind this result? Why does the dynamic
equation, written in the relativistically covariant form, become to be
nonrelativistic, if it contains an absolute vector? To answer this question,
we consider an example of a charged classical particle, moving in the given
electromagnetic field $F^{ik}$.

Dynamic equations for the relativistic particle may be written in the
noncovariant form%
\begin{equation}
\frac{d}{dt}\frac{m\dot{x}^{\mu }}{\sqrt{1-\frac{\mathbf{\dot{x}}^{2}}{c^{2}}%
}}=\frac{e}{c}F^{\mu 0}+\frac{e}{c}F^{\mu \nu }g_{\nu \beta }\dot{x}^{\beta
},\qquad \mu =1,2,3,\qquad \mathbf{\dot{x}}\equiv \frac{d\mathbf{x}}{dt}
\label{a9.1}
\end{equation}%
and in the relativistically covariant form%
\begin{equation}
m\frac{d^{2}x^{k}}{d\tau ^{2}}=\frac{e}{c}F^{kl}g_{ls}\frac{dx^{s}}{d\tau }%
,\qquad k=0,1,2,3  \label{a9.2}
\end{equation}%
where $\tau $ is the proper time, $e$, $m$ are respectively the particle
charge and the particle mass.

If the particle is nonrelativistic the dynamic equations are written in the
noncovariant form%
\begin{equation}
m\frac{d^{2}x^{\mu }}{dt^{2}}=\frac{e}{c}F^{\mu 0}+\frac{e}{c}F^{\mu \nu
}g_{\nu \beta }\frac{dx^{\beta }}{dt},\qquad \mu =1,2,3  \label{a9.3}
\end{equation}%
Can the dynamic equations (\ref{a9.3}) for the nonrelativistic particle be
written in the relativistically covariant form? The answer is yes, although
most researchers believe that it is impossible. In the relativistically
covariant form the dynamic equations (\ref{a9.3}) have the form 
\begin{equation}
m\frac{d}{d\tau }\left[ \left( l_{k}\dot{x}^{k}\right) ^{-1}\dot{x}^{i}-{%
\frac{1}{2}}g^{ik}l_{k}\left( l_{j}\dot{x}^{j}\right) ^{-2}\dot{x}^{s}g_{sl}%
\dot{x}^{l}\right] =\frac{e}{c}F^{il}g_{lk}\dot{x}^{k};\qquad i=0,1,2,3
\label{a9.4}
\end{equation}%
where $\dot{x}^{k}\equiv dx^{k}/d\tau $. The quantity $l_{k}$, $k=0,1,2,3$
is a constant timelike unit 4-vector 
\begin{equation}
g^{ik}l_{i}l_{k}=1;  \label{a9.5}
\end{equation}%
Using the special choice of $l_{k}=\left\{ c,0,0,0\right\} $ and
substituting it in (\ref{a9.4}), it is easy to verify, that we obtain the
dynamic equations (\ref{a9.3}) for $i=1,2,3$. For $i=0$ we obtain dynamic
equation, which is a corollary of (\ref{a9.3}).

As far as dynamic equations for both relativistic and nonrelativistic
particles can be written in the noncovariant form and in the
relativistically covariant one, it is clear that the difference between the
relativistic and nonrelativistic dynamic equations is not connected with
form of dynamic equations. There is anything else, which distinguishes the
relativistic conception from the nonrelativistic one.

It is well known that the difference lies in different space-time
conceptions. In the Newtonian conception there is an absolute simultaneity
and there are two invariant quantities: absolute time $t$ and absolute space
distance $r$, whereas in the relativistic space-time conception there exists
only one absolute quantity: the space-time interval $s=\sqrt{c^{2}t^{2}-r^{2}%
}$. The Newtonian space-time $\mathcal{S}_{\mathrm{N}}$ has seven-parametric
continuous group of motion, whereas the Minkowski space-time $\mathcal{S}_{%
\mathrm{M}}$ has ten-parametric continuous group of motion. Besides, the
Newtonian space-time $\mathcal{S}_{\mathrm{N}}$ may be considered to be the
Minkowski space-time $\mathcal{S}_{\mathrm{M}}$ with additional geometric
structure $\mathcal{L}$, given in it. In other words, $\mathcal{S}_{\mathrm{N%
}}=\mathcal{S}_{\mathrm{M}}\wedge \mathcal{L}$. The additional structure $%
\mathcal{L}$ is a specific timelike direction in $\mathcal{S}_{\mathrm{M}}$,
described by the constant timelike vector $l_{k}$. Any hyperplane,
orthogonal to $l_{k}$, is a set of absolutely simultaneous events.
Introduction of $\mathcal{L}$ admits one to construct two invariants in $%
\mathcal{S}_{\mathrm{M}}\wedge \mathcal{L}$ 
\begin{equation}
t=l_{k}x^{k},\qquad r=\sqrt{x^{k}x_{k}+\left( l_{k}x^{k}\right) ^{2}}
\label{a9.6}
\end{equation}%
for any vector $x^{k}$, whereas in $\mathcal{S}_{\mathrm{M}}$ we have only
one invariant $s=\sqrt{x^{k}x_{k}}$. Construction of this invariant does not
contain a reference to $\mathcal{L}$.

The Newtonian space-time $\mathcal{S}_{\mathrm{N}}$ considered as $\mathcal{S%
}_{\mathrm{M}}\wedge \mathcal{L}$ admits only such motions of $\mathcal{S}_{%
\mathrm{M}}$, which transform vector $l_{k}$ into the same vector $l_{k}$
and do not violate the structure $\mathcal{L}$. The condition of the
structure $\mathcal{L}$ conservation at the space-time motion reduces the
ten-parametric group of motion of $\mathcal{S}_{\mathrm{M}}$ to
seven-parametric group of motion of $\mathcal{S}_{\mathrm{M}}\wedge \mathcal{%
L}$. In general, at the relativistically covariant description the absolute
objects, introduced by Anderson \cite{A67}, may be considered as the
quantities, describing additional structures in $\mathcal{S}_{\mathrm{M}}$.
It means, that any system of dynamic equations may be written in the
relativistically covariant form, provided the proper absolute objects
(additional structures) are introduced. Thus, to determine, whether the
dynamic equations are compatible with the principles of relativity, we may
write them in the relativistically covariant form and determine whether or
not they contain absolute objects and what are properties of these absolute
objects. If the dynamic equations contain the constant timelike vector $%
l_{k} $, we have nonrelativistic dynamic system, because $l_{k}$ describes
the additional space-time structure, characteristic for the Newtonian
space-time $\mathcal{S}_{\mathrm{N}}$ represented as $\mathcal{S}_{\mathrm{M}%
}\wedge \mathcal{L}$.

Such an approach is convenient in the sense, that it does not contain a
reference to the coordinate system, which is simply a method of description.
Relativistic character of dynamic equation is connected directly with
absence of additional space-time structures in $\mathcal{S}_{\mathrm{M}}$,
but not with the relativistically covariant form of the dynamic equations,
because any dynamic equations can be always written in the relativistically
covariant form, provided the proper geometrical structure is introduced in $%
\mathcal{S}_{\mathrm{M}}$. The relativistically covariant representation of
dynamic equations is \textit{necessary only for a discovery of additional
geometrical structures} in the Minkowski space-time $\mathcal{S}_{\mathrm{M}%
} $. \textit{Additional geometeric structures are primary}, whereas the
relativistic covariance is secondary, because it admits one only to discover
these structures if they takes place. The relativistic covariance in itself
is indifferent with respect to the relativistic invariance of the dynamic
equations. The additional structure is formally present in the equations (%
\ref{a9.4}) and described by the formal parameters $l_{k}$. When we use
substitution $l_{k}=\left\{ c,0,0,0\right\} $ in (\ref{a9.4}), we obtain (%
\ref{a9.3}), where the structure $\mathcal{L}$ is formally absent, because
the formal parameters $l_{k}$ of $\mathcal{L}$ are absent.

Thus, the relativistic invariance of the dynamic equations is connected with
existence and properties of additional structures in $\mathcal{S}_{\mathrm{M}%
}$, whereas the relativistic covariance is only \textit{a method of discovery%
} of these structures.

The relativistically covariant dynamic equation is relativistic, provided it
does not contain a reference to some additional structure. However, such a
formulation is unreliable, because the reservation of a reference to
additional structure may be omitted by mistake. In this case the
relativistic character of dynamic equations appears to be connected with the
relativistic covariance of these equations, but not with the additional
structure $\mathcal{L}$ in $\mathcal{S}_{\mathrm{M}}$. It is this case that
takes place in reality. As a result we have an associative mistake, when the
relativistic invariance is associated with the relativistic covariance,
although in reality the relativistic invariance is associated with an
absence of additional geometrical structures in $\mathcal{S}_{\mathrm{M}}$.
Relativistic covariant form of dynamic equations is only a condition, when
existence of additional structure in $\mathcal{S}_{\mathrm{M}}$ can be
discovered.

Thus, the internal degrees of freedom of the Dirac particle has not been
discovered theoretically, because they cannot be obtained experimentally.
All precise experiments with such a Dirac particle as electron are
nonrelativistic (correction to the spectrum of hydrogen) and internal rigid
degrees of freedom give only a negligible correction. As far as the internal
degrees of freedom remained unknown, one cannot obtain experimentally
nonrelativistic character of their description. Unfortunately, the quantum
principles do not work in the relativistic region, and the conventional
quantum theory cannot obtain those results, which has not been obtained
experimentally, because it needs additional hypotheses, having experimental
basis. The dynamical methods are free of this defect. They work
independently of experimental data and additional hypotheses.

\renewcommand{\theequation}{\Alph{section}.\arabic{equation}} %
\renewcommand{\thesection}{\Alph{section}} \setcounter{section}{0} %
\centerline{\Large \bf Mathematical Appendix}

\section{Transformation of the action for the statistical ensemble}

Let us transform the action 
\begin{equation}
\mathcal{E}\left[ \mathcal{S}_{\mathrm{st}}\right] :\qquad \mathcal{A}_{%
\mathcal{E}\left[ \mathcal{S}_{\mathrm{st}}\right] }\left[ \mathbf{x},%
\mathbf{u}_{\mathrm{st}}\right] \mathbf{=}\int \left\{ \frac{m\mathbf{\dot{x}%
}^{2}}{2}-\frac{e}{c}A_{0}-\frac{e}{c}\mathbf{A}\frac{d\mathbf{x}}{dt}+\frac{%
m\mathbf{u}_{\mathrm{st}}^{2}}{2}-\frac{\hbar }{2}\mathbf{\nabla u}_{\mathrm{%
st}}\right\} dtd\mathbf{\xi }  \label{g1.19}
\end{equation}%
for the statistical ensemble of stochastic particles, moving in the given
electromagnetic field $A=\left\{ A_{0},\mathbf{A}\right\} =\left\{
A_{0},A_{1},A_{2},A_{3}\right\} $. Here $\mathbf{x}=\mathbf{x}\left( t,%
\mathbf{\xi }\right) $, $\mathbf{u}_{\mathrm{st}}=\mathbf{u}_{\mathrm{st}%
}\left( t,\mathbf{x}\right) $ are dependent dynamic variables, and $\mathbf{%
\nabla =}\left\{ \partial _{1},\partial _{2},\partial _{3}\right\} \mathbf{=}%
\left\{ \frac{\partial }{\partial x^{1}},\frac{\partial }{\partial x^{2}},%
\frac{\partial }{\partial x^{3}}\right\} $. The variable $\mathbf{x}$
describes the regular component of the stochastic particle motion. The
dynamic variable $\mathbf{u}_{\mathrm{st}}$ is a function of $t,\mathbf{x}$
and depends on $\mathbf{\xi }$ via $\mathbf{x}$. The quantity $\mathbf{u}_{%
\mathrm{st}}$ may be regarded as the mean velocity of the stochastic
component, whereas $\mathbf{x}=\mathbf{x}\left( t,\mathbf{\xi }\right) $
describes the regular component of the particle motion. The last term in (%
\ref{g1.19}) describes influence of the stochasticity on the evolution of
the regular component.

To eliminate variable $\mathbf{u}_{\mathrm{st}}$, we should to solve dynamic
equations $\delta \mathcal{A}/\mathcal{\delta }\mathbf{u}_{\mathrm{st}}=0$
with respect to $\mathbf{u}_{\mathrm{st}}$. As far as $\mathbf{u}_{\mathrm{st%
}}$ is a function of $t,\mathbf{x}$, we should go to independent variables $%
t,\mathbf{x}$ in the action (\ref{g1.19}). We obtain instead of (\ref{g1.19}%
) 
\begin{equation}
\mathcal{A}_{\mathcal{E}\left[ \mathcal{S}_{\mathrm{st}}\right] }\left[ 
\mathbf{\xi },\mathbf{u}_{\mathrm{st}}\right] \mathbf{=}\int \left\{ \frac{m%
\mathbf{\dot{x}}^{2}}{2}-\frac{e}{c}A_{0}-\frac{e}{c}\mathbf{A}\frac{d%
\mathbf{x}}{dt}+\frac{m\mathbf{u}_{\mathrm{st}}^{2}}{2}-\frac{\hbar }{2}%
\mathbf{\nabla u}_{\mathrm{st}}\right\} \rho \left( t,\mathbf{x}\right) dtd%
\mathbf{x}  \label{g1.20}
\end{equation}
where $\mathbf{\ \xi }$, $\mathbf{u}_{\mathrm{st}}$ are dependent variables,
whereas $t,\mathbf{x}$ are independent variables. Here $\rho $ and $\mathbf{%
\dot{x}}=\mathbf{u}$ are functions of $\mathbf{\xi }$, defined by the
relations 
\begin{equation}
\rho =\frac{\partial \left( \xi _{1},\xi _{2},\xi _{3}\right) }{\partial
\left( x^{1},x^{2},x^{3}\right) },\qquad \mathbf{\dot{x}}\equiv \mathbf{u}%
\equiv \frac{\partial \left( \mathbf{x,}\xi _{1},\xi _{2},\xi _{3}\right) }{%
\partial \left( t,\xi _{1},\xi _{2},\xi _{3}\right) }=\frac{1}{\rho }\frac{%
\partial \left( \mathbf{x,}\xi _{1},\xi _{2},\xi _{3}\right) }{\partial
\left( t,x^{1},x^{2},x^{3}\right) },  \label{g1.21}
\end{equation}
Variation of (\ref{g1.20}) with respect $\mathbf{u}_{\mathrm{st}}$ gives 
\begin{equation}
\frac{\delta \mathcal{A}_{\mathcal{E}\left[ \mathcal{S}_{\mathrm{st}}\right]
}}{\mathcal{\delta }\mathbf{u}_{\mathrm{st}}}=m\mathbf{u}_{\mathrm{st}}\rho +%
\frac{\hbar }{2}\mathbf{\nabla }\rho =0  \label{g1.22}
\end{equation}
Resolving the equation (\ref{g1.22}) with respect to $\mathbf{u}_{\mathrm{st}%
}$ in the form 
\begin{equation}
\mathbf{u}_{\mathrm{st}}=-\frac{\hbar }{2m}\mathbf{\nabla }\ln \rho ,
\label{g1.23}
\end{equation}
we obtain instead of (\ref{g1.20}) 
\begin{equation}
\mathcal{A}_{\mathcal{E}\left[ \mathcal{S}_{\mathrm{st}}\right] }\left[ 
\mathbf{\xi }\right] \mathbf{=}\int \left\{ \frac{m}{2}\left( \frac{d\mathbf{%
x}}{dt}\right) ^{2}-\frac{e}{c}A_{0}-\frac{e}{c}\mathbf{A}\frac{d\mathbf{x}}{%
dt}-\frac{\hbar ^{2}}{8m}\frac{\left( \mathbf{\nabla }\rho \right) ^{2}}{%
\rho ^{2}}\right\} \rho dtd\mathbf{x}  \label{g1.24}
\end{equation}
where $\rho $ and $\frac{d\mathbf{x}}{dt}$ are functions of space-time
derivatives of $\mathbf{\xi }=\left\{ \xi _{1},\xi _{2},\xi _{3}\right\} $,
determined by the relations (\ref{g1.21}). The action (\ref{g1.24})
describes some ideal charged fluid with the internal energy per unit mass 
\begin{equation}
U\left( \rho ,\mathbf{\nabla }\rho \right) =\frac{\hbar ^{2}}{8m}\frac{%
\left( \mathbf{\nabla }\rho \right) ^{2}}{\rho ^{2}}  \label{g1.25}
\end{equation}

Let us introduce new dependent variables $j=\left\{ \rho ,\rho \mathbf{u}%
\right\} =\left\{ j^{k}\right\} ,\;\;k=0,1,2,3$ by means of relations (\ref%
{g1.21}). From formal viewpoint it is convenient to represent the
hydrodynamic variables $j=\left\{ \rho ,\rho \mathbf{u}\right\} =\left\{
j^{k}\right\} $, $k=0,1,2,3$ in the form 
\begin{equation}
j^{k}=\frac{\partial \left( x^{k}\mathbf{,}\xi _{1},\xi _{2},\xi _{3}\right) 
}{\partial \left( x^{0},x^{1},x^{2},x^{3}\right) }=\frac{\partial J}{%
\partial \xi _{0,k}},\qquad k=0,1,2,3  \label{g1.26}
\end{equation}
where the Jacobian 
\begin{equation}
J=\frac{\partial \left( \xi _{0}\mathbf{,}\xi _{1},\xi _{2},\xi _{3}\right) 
}{\partial \left( x^{0},x^{1},x^{2},x^{3}\right) }=\det \left\vert
\left\vert \xi _{i,k}\right\vert \right\vert ,\qquad \xi _{l,k}\equiv
\partial _{k}\xi _{l},\qquad l,k=0,1,2,3  \label{g1.27}
\end{equation}
is considered to be a function of variables $\xi _{l,k}\equiv \partial
_{k}\xi _{l},\;l,k=0,1,2,3$. The variable $\xi _{0}$ is the new dependent
variable (temporal Lagrangian coordinate), which appears to be fictitious.

We introduce new dynamic variables by the Lagrange multipliers $p=\left\{
p_{k}\right\} ,\;\;k=0,1,2,3$, and obtain instead of (\ref{g1.24}) 
\begin{equation}
\mathcal{A}_{\mathcal{E}\left[ \mathcal{S}_{\mathrm{st}}\right] }\left[ \xi 
\mathbf{,}j,p\right] \mathbf{=}\int \left\{ \frac{m}{2\rho }j^{\alpha
}j^{\alpha }-\frac{e}{c}A_{0}\rho -\frac{e}{c}A_{\alpha }j^{\alpha
}-p_{k}\left( j^{k}-\frac{\partial J}{\partial \xi _{0,k}}\right) -\frac{%
\hbar ^{2}}{8m}\frac{\left( \mathbf{\nabla }\rho \right) ^{2}}{\rho }%
\right\} d^{4}x  \label{g1.28}
\end{equation}%
where $\xi =\left\{ \xi _{k}\right\} $,$\;\;k=0,1,2,3$.

Variation of the action (\ref{g1.28}) with respect to $\xi _{l}$ leads to
the dynamic equations 
\begin{equation}
\frac{\delta \mathcal{A}_{\mathcal{E}\left[ \mathcal{S}_{\mathrm{st}}\right]
}}{\mathcal{\delta \xi }_{l}}=\partial _{s}\left( p_{k}\frac{\partial ^{2}J}{%
\partial \xi _{0,k}\partial \xi _{l,s}}\right) =0,\qquad l=0,1,2,3
\label{g1.29}
\end{equation}
As far as the variable $\xi _{0}$ is fictitious, there are only three
independent equations among four equations (\ref{g1.29}).

Using identities 
\begin{equation}
\frac{\partial ^{2}J}{\partial \xi _{0,k}\partial \xi _{l,s}}\equiv
J^{-1}\left( \frac{\partial J}{\partial \xi _{0,k}}\frac{\partial J}{%
\partial \xi _{l,s}}-\frac{\partial J}{\partial \xi _{0,s}}\frac{\partial J}{%
\partial \xi _{l,k}}\right)  \label{A.9}
\end{equation}
\begin{equation}
\frac{\partial J}{\partial \xi _{i,l}}\xi _{k,l}\equiv J\delta
_{k}^{i},\qquad \partial _{l}\frac{\partial ^{2}J}{\partial \xi
_{0,k}\partial \xi _{i,l}}\equiv 0  \label{A.10}
\end{equation}
and designations (\ref{g1.26}), we can eliminate the variables $\mathbf{\xi }
$ from the equations (\ref{g1.29}). We obtain 
\begin{equation}
j^{k}\partial _{l}p_{k}-j^{k}\partial _{k}p_{l}=0,\qquad l=0,1,2,3
\label{A.10a}
\end{equation}

Variation of (\ref{g1.28}) with respect to $j^{\beta }$ and $j^{0}=\rho $
gives respectively 
\begin{equation}
p_{\beta }=m\frac{j^{\beta }}{\rho }-\frac{e}{c}A_{\beta },\qquad \beta
=1,2,3  \label{A.17}
\end{equation}
\begin{equation}
p_{0}=-\frac{m}{2\rho ^{2}}j^{\alpha }j^{\alpha }-\frac{e}{c}A_{0}+\frac{%
\hbar ^{2}}{8m}\left( 2\frac{\mathbf{\nabla }^{2}\rho }{\rho }-\frac{\left( 
\mathbf{\nabla }\rho \right) ^{2}}{\rho ^{2}}\right)  \label{A.10b}
\end{equation}
Eliminating $p_{k}$ from the equations (\ref{A.10a}) by means of relations (%
\ref{A.17}), (\ref{A.10b}), we obtain hydrodynamic equations for the ideal
charged fluid in the conventional form 
\begin{equation}
\left( \partial _{0}+v^{\alpha }\partial _{\alpha }\right) v^{\mu }=\frac{e}{%
mc}F_{\mu 0}+\frac{e}{mc}F_{\mu \alpha }v^{\alpha }-\frac{1}{m\rho }\partial
_{\mu }p,\qquad \mu =1,2,3  \label{A.10c}
\end{equation}
where the pressure $p$ and the electromagnetic field $F_{ik}$ are defined by
the relations 
\begin{equation}
p=\frac{\hbar ^{2}}{8m}\left( \frac{\left( \mathbf{\nabla }\rho \right) ^{2}%
}{\rho ^{2}}-2\frac{\mathbf{\nabla }^{2}\rho }{\rho }\right) ,\qquad
F_{ik}=\partial _{k}A_{i}-\partial _{i}A_{k},\qquad i,k=0,1,2,3
\label{A.10d}
\end{equation}

The wave function is constructed of potentials. The equations (\ref{A.10c})
does not contain potentials $\mathbf{\xi }$ and $A_{k}$, and they cannot be
used for description of the fluid in terms of the wave function. To
construct a description in terms of the wave function, we should not to
eliminate potentials $\mathbf{\xi }$ from the equations (\ref{g1.29}).
Instead, we should integrate them. The dynamic equations (\ref{g1.29}) may
be considered to be linear partial differential equations with respect to
variables $p_{k}$. They can be solved in the form 
\begin{equation}
p_{k}=b\left( \partial _{k}\varphi +g^{\alpha }\left( \mathbf{\xi }\right)
\partial _{k}\xi _{\alpha }\right) ,\qquad k=0,1,2,3  \label{g1.30}
\end{equation}%
where $g^{\alpha }\left( \mathbf{\xi }\right) ,\;\;\alpha =1,2,3$ are
arbitrary functions of the argument $\mathbf{\xi }=\left\{ \xi _{1},\xi
_{2},\xi _{3}\right\} $, $b$ is an arbitrary real constant, and $\varphi $
is the variable $\xi _{0}$, which ceases to be fictitious.

One can test by the direct substitution that the relation (\ref{g1.30}) is
the general solution of linear equations (\ref{g1.29}). Indeed, using (\ref%
{A.9}) and the second identity (\ref{A.10}), the equations (\ref{g1.29}) may
be written in the form 
\begin{equation}
\frac{\partial ^{2}J}{\partial \xi _{0,k}\partial \xi _{l,s}}\partial
_{s}p_{k}=J^{-1}\left( \frac{\partial J}{\partial \xi _{0,k}}\frac{\partial J%
}{\partial \xi _{l,s}}-\frac{\partial J}{\partial \xi _{0,s}}\frac{\partial J%
}{\partial \xi _{l,k}}\right) \partial _{s}p_{k}=0  \label{g1.31}
\end{equation}
Substituting (\ref{g1.30}) in (\ref{g1.31}) and taking into account
antisymmetry of the bracket in (\ref{g1.31}) with respect to indices $k$ and 
$s$, we obtain 
\begin{equation}
J^{-1}\left( \frac{\partial J}{\partial \xi _{0,k}}\frac{\partial J}{%
\partial \xi _{l,s}}-\frac{\partial J}{\partial \xi _{0,s}}\frac{\partial J}{%
\partial \xi _{l,k}}\right) \frac{\partial g^{\alpha }\left( \mathbf{\xi }%
\right) }{\partial \xi _{\mu }}\xi _{\mu ,s}\xi _{\alpha ,k}=0  \label{g1.32}
\end{equation}
The relation (\ref{g1.32}) is the valid equality, as it follows from the
first identity (\ref{A.10}).

Let us substitute (\ref{g1.30}) in the action (\ref{g1.28}). Taking into
account the first identity (\ref{A.10}) and omitting the term 
\begin{equation*}
\frac{\partial J}{\partial \xi _{0,k}}\partial _{k}\varphi =\frac{\partial
\left( \varphi ,\xi _{1},\xi _{2},\xi _{3}\right) }{\partial \left(
x^{0},x^{1},x^{2},x^{3}\right) }
\end{equation*}
which does not contribute to the dynamic equation, we obtain 
\begin{equation}
\mathcal{E}\left[ \mathcal{S}_{\mathrm{st}}\right] :\qquad \mathcal{A}_{%
\mathcal{E}\left[ \mathcal{S}_{\mathrm{st}}\right] }\left[ \varphi ,\mathbf{%
\xi },j\right] =\int \left\{ \frac{m}{2}\frac{j^{\alpha }j^{\alpha }}{j^{0}}-%
\frac{e}{c}A_{k}j^{k}-j^{k}p_{k}-\frac{\hbar ^{2}}{8m}\frac{\left( \mathbf{%
\nabla }\rho \right) ^{2}}{\rho }\right\} d^{4}x\mathbf{,}  \label{A.12}
\end{equation}
Here quantities $p_{k}$ are determined by the relations (\ref{g1.30}).

The action in the form (\ref{A.12}) is remarkable in the sense, that it
contains information on initial values of the fluid velocities $\mathbf{v}=%
\mathbf{j}/\rho $. Dynamic equations, generated by the action (\ref{A.12}),
are partial differential equations, and one needs to give initial values for
variables $\varphi ,\mathbf{\xi }$\textbf{.} But initial values for
variables $\varphi ,\mathbf{\xi }$ determine only labelling of the fluid
particles, and they may be chosen universal. For instance, we may choose for
all fluid flows 
\begin{equation}
\varphi \left( 0,\mathbf{x}\right) =\varphi _{\mathrm{in}}\left( \mathbf{x}%
\right) =0,\qquad \mathbf{\xi }\left( 0,\mathbf{x}\right) =\mathbf{\xi }_{%
\mathrm{in}}\left( \mathbf{x}\right) =\mathbf{x}  \label{A.12b}
\end{equation}
Then the functions $\mathbf{g}\left( \mathbf{\xi }\right) $ are determined
by the initial values of the velocity $\mathbf{v}\left( 0,\mathbf{x}\right) =%
\mathbf{v}_{\mathrm{in}}\left( \mathbf{x}\right) $ in the form \cite{R99} 
\begin{equation}
\mathbf{g}\left( \mathbf{\xi }\right) =\mathbf{v}_{\mathrm{in}}\left( 
\mathbf{\xi }\right)  \label{A.12c}
\end{equation}
The initial value $\rho \left( 0,\mathbf{x}\right) =\rho _{\mathrm{in}%
}\left( \mathbf{x}\right) $ of the density $\rho $ may be also included in
the action (\ref{A.12}). It is necessary only to redefine the connection
between the quantities $j^{k}$ and $\mathbf{\xi }$\textbf{, }substituting
the relations (\ref{g1.26}) by the relations \cite{R99} 
\begin{equation}
j^{k}=\rho _{0}\left( \mathbf{\xi }\right) \frac{\partial \left( x^{k}%
\mathbf{,}\xi _{1},\xi _{2},\xi _{3}\right) }{\partial \left(
x^{0},x^{1},x^{2},x^{3}\right) },\qquad k=0,1,2,3  \label{A.12d}
\end{equation}
where $\rho _{0}\left( \mathbf{\xi }\right) $ is an arbitrary function of $%
\mathbf{\xi }$. At the initial conditions (\ref{A.12c}) this arbitrary
function is to be chosen in the form 
\begin{equation*}
\rho _{0}\left( \mathbf{x}\right) =\rho _{\mathrm{in}}\left( \mathbf{x}%
\right) =\rho \left( 0,\mathbf{x}\right)
\end{equation*}

Now we eliminate the variables $\mathbf{j}=\left\{ j^{1},j^{2},j^{3}\right\} 
$ from the action (\ref{A.12}), using relation (\ref{A.17}). We obtain 
\begin{equation}
\mathcal{A}_{\mathcal{E}\left[ \mathcal{S}_{\mathrm{st}}\right] }\left[ \rho
,\varphi ,\mathbf{\xi }\right] =\int \left\{ -p_{0}-\frac{e}{c}A_{0}-\frac{%
\left( p_{\beta }+\frac{e}{c}A_{\beta }\right) \left( p_{\beta }+\frac{e}{c}%
A_{\beta }\right) }{2m}-\frac{\hbar ^{2}}{8m}\frac{\left( \mathbf{\nabla }%
\rho \right) ^{2}}{\rho ^{2}}\right\} \rho d^{4}x\mathbf{,}  \label{A.18}
\end{equation}%
where the quantities $p_{k}$, $k=0,1,2,3$ are determined by the relation (%
\ref{g1.30}).

Instead of dependent variables $\rho ,\varphi ,\mathbf{\xi }$ we introduce
the $n$-component complex function $\psi =\{\psi _{\alpha }\},\;\;\alpha
=1,2,\ldots ,n$, which is defined by the relations \cite{R99} 
\begin{equation}
\psi _{\alpha }=\sqrt{\rho }e^{i\varphi }u_{\alpha }(\mathbf{\xi }),\qquad
\psi _{\alpha }^{\ast }=\sqrt{\rho }e^{-i\varphi }u_{\alpha }^{\ast }(%
\mathbf{\xi }),\qquad \alpha =1,2,\ldots ,n,  \label{A.19}
\end{equation}
\begin{equation}
\psi ^{\ast }\psi \equiv \sum_{\alpha =1}^{n}\psi _{\alpha }^{\ast }\psi
_{\alpha },  \label{A.20}
\end{equation}
where (*) means the complex conjugate. The quantities $u_{\alpha }(\mathbf{%
\xi })$, $\alpha =1,2,\ldots ,n$ are functions of only variables $\mathbf{%
\xi }$, and satisfy the relations 
\begin{equation}
-\frac{i}{2}\sum_{\alpha =1}^{n}\left( u_{\alpha }^{\ast }\frac{\partial
u_{\alpha }}{\partial \xi _{\beta }}-\frac{\partial u_{\alpha }^{\ast }}{%
\partial \xi _{\beta }}u_{\alpha }\right) =g^{\beta }(\mathbf{\xi }),\qquad
\beta =1,2,3,\qquad \sum_{\alpha =1}^{n}u_{\alpha }^{\ast }u_{\alpha }=1.
\label{A.21}
\end{equation}
The number $n$ is such a natural number that the equations (\ref{A.21})
admit a solution. In general, $n$ depends on the form of the arbitrary
integration functions $\mathbf{g}=\{g^{\beta }(\mathbf{\xi })\}$, $\beta
=1,2,3$. The functions $\mathbf{g}$ determine vorticity of the fluid flow.
If $\mathbf{g}=0$, equations (\ref{A.21}) have the solution $u_{1}=1$, $%
u_{\alpha }=0$, \ $\alpha =2,3,...n$. In this case the function $\psi $ may
have one component, and the fluid flow is irrotational.

In the general case it is easy to verify that 
\begin{equation}
\rho =\psi ^{\ast }\psi ,\qquad \rho p_{0}\left( \varphi ,\mathbf{\xi }%
\right) =-\frac{ib}{2}(\psi ^{\ast }\partial _{0}\psi -\partial _{0}\psi
^{\ast }\cdot \psi )  \label{s5.6}
\end{equation}%
\begin{equation}
\rho p_{\alpha }\left( \varphi ,\mathbf{\xi }\right) =-\frac{ib}{2}(\psi
^{\ast }\partial _{\alpha }\psi -\partial _{\alpha }\psi ^{\ast }\cdot \psi
),\qquad \alpha =1,2,3,  \label{s5.7}
\end{equation}%
The variational problem with the action (\ref{A.12}) appears to be
equivalent to the variational problem with the action functional 
\begin{equation*}
\mathcal{A}_{\mathcal{E}\left[ \mathcal{S}_{\mathrm{st}}\right] }[\psi ,\psi
^{\ast }]=\int \left\{ \frac{ib}{2}(\psi ^{\ast }\partial _{0}\psi -\partial
_{0}\psi ^{\ast }\cdot \psi )-\frac{e}{c}A_{0}\rho \right.
\end{equation*}%
\begin{equation}
\left. -\frac{\rho }{2m}\left( -\frac{ib}{2\rho }(\psi ^{\ast }\mathbf{%
\nabla }\psi -\mathbf{\nabla }\psi ^{\ast }\cdot \psi )+\frac{e}{c}\mathbf{A}%
\right) ^{2}-\frac{\hbar ^{2}}{8m}\frac{\left( \mathbf{\nabla }\rho \right)
^{2}}{\rho }\right\} \mathrm{d}^{4}x  \label{s5.8}
\end{equation}%
or%
\begin{equation*}
\mathcal{A}_{\mathcal{E}\left[ \mathcal{S}_{\mathrm{st}}\right] }[\psi ,\psi
^{\ast }]=\int \left\{ \frac{ib}{2}(\psi ^{\ast }\partial _{0}\psi -\partial
_{0}\psi ^{\ast }\cdot \psi )-\frac{e}{c}A_{0}+\frac{b^{2}}{8m\rho }(\psi
^{\ast }\mathbf{\nabla }\psi -\mathbf{\nabla }\psi ^{\ast }\cdot \psi
)^{2}\right.
\end{equation*}
\begin{equation}
\left. +\frac{ibe}{2mc}\mathbf{A}(\psi ^{\ast }\mathbf{\nabla }\psi -\mathbf{%
\nabla }\psi ^{\ast }\cdot \psi )-\frac{\hbar ^{2}}{8m}\frac{\left( \mathbf{%
\nabla }\rho \right) ^{2}}{\rho }-\frac{\rho }{2m}\left( \frac{e}{c}\mathbf{A%
}\right) ^{2}\right\} \mathrm{d}^{4}x  \label{s5.9}
\end{equation}

For the two-component function $\psi $ \ ($n=2$) the following identity
takes place 
\begin{equation}
(\mathbf{\nabla }\rho )^{2}-(\psi ^{\ast }\mathbf{\nabla }\psi -\mathbf{%
\nabla }\psi ^{\ast }\cdot \psi )^{2}\equiv 4\rho \mathbf{\nabla }\psi
^{\ast }\mathbf{\nabla }\psi -\rho ^{2}\sum\limits_{\alpha =1}^{\alpha
=3}\left( \mathbf{\nabla }s_{\alpha }\right) ^{2},  \label{s5.30}
\end{equation}%
\begin{equation}
\rho \equiv \psi ^{\ast }\psi ,\qquad \mathbf{s}\equiv \frac{\psi ^{\ast }%
\mathbf{\sigma }\psi }{\rho },\qquad \mathbf{\sigma }=\{\sigma _{\alpha
}\},\qquad \alpha =1,2,3,  \label{s5.30a}
\end{equation}%
where $\sigma _{\alpha }$ are the Pauli matrices. In virtue of the identity (%
\ref{s5.30}) the action (\ref{s5.8}) reduces to the form%
\begin{equation*}
\mathcal{A}_{\mathcal{E}\left[ \mathcal{S}_{\mathrm{st}}\right] }[\psi ,\psi
^{\ast }]
\end{equation*}%
\begin{equation*}
=\int \left\{ \frac{ib}{2}\left( \psi ^{\ast }\partial _{0}\psi -\partial
_{0}\psi ^{\ast }\cdot \psi \right) -\frac{e}{c}A_{0}-\frac{1}{2m}\left( -ib%
\mathbf{\nabla }\psi ^{\ast }-\frac{e}{c}\mathbf{A}\psi ^{\ast }\right)
\left( ib\mathbf{\nabla }\psi -\frac{e}{c}\mathbf{A}\psi \right) \right.
\end{equation*}
\begin{equation}
\left. +\frac{b^{2}-\hbar ^{2}}{8\rho m}(\mathbf{\nabla }\rho )^{2}+\frac{%
b^{2}}{8m}\sum\limits_{\alpha =1}^{\alpha =3}(\mathbf{\nabla }s_{\alpha
})^{2}\rho \right\} \mathrm{d}^{4}x,  \label{s5.32}
\end{equation}%
where $\mathbf{s}$ and $\rho $ are defined by the relations (\ref{s5.30a}).
One should expect, that the two-component wave function describes the
general case, because the number of real components of the two-component
wave function coincides with the number of hydrodynamic variables $\left\{
\rho ,\mathbf{j}\right\} $. But this statement is not yet proved.

In the case of irrotational flow, when the two-component function $\psi $
has linear dependent components, for instance $\psi =\left\{ \psi
_{1},0\right\} $, the 3-vector $\mathbf{s}=$const, and the term containing
3-vector $\mathbf{s}$ vanishes. In the special case, when the
electromagnetic potentials $A_{k}=0$, the action (\ref{s5.32}) for $\mathcal{%
E}\left[ \mathcal{S}_{\mathrm{st}}\right] $ coincides with the action (\ref%
{a1.10}) for $\mathcal{S}_{\mathrm{S}}$.

Finally, if we choose the arbitrary constant $b$ in the form $b=\hbar $ and
set $A_{k}=0$, we obtain the action (\ref{a1.2}) for the Schr\"{o}dinger
particle.

\end{document}